\documentclass[preprint,notoc]{JHEP3}
\usepackage{epsfig}
\def\fa{f_a}

\def\dsl{\partial \llap/}
\def\sr2{\sqrt{2}}

\def\to{\rightarrow}

\def\bi{\begin{itemize}}
\def\ei{\end{itemize}}

\def\c1p{C1^\prime}
\def\ta{\tilde a}
\def\tG{\widetilde G}

\def\tu{\tilde u}

\def\ta{\tilde a}

\def\tst{\tilde t}

\def\tg{\tilde g}

\def\tq{\tilde q}

\def\tz{\widetilde Z}

\def\sigv{\langle \sigma v \rangle}
\def\To{\Rightarrow}
\def\alt{\lesssim}
\def\agt{\gtrsim}
\def\be{\begin{equation}}  
\def\ee{\end{equation}}  
\def\bea{\begin{eqnarray}}  
\def\eea{\end{eqnarray}}

\def\sigv{\langle \sigma v \rangle}
\def\To{\Rightarrow}
\newcommand\drv[2]{\frac{\partial #1}{\partial #2}}
\newcommand\Drv[2]{\frac{d #1}{d #2}}

\newcommand\sjp[3]{{\it Sov.\ J.\ Nucl.\ }{\bf #1} (#2) #3}

\def\Isajet{{\sc Isajet}}

\preprint{\vbox{OU-HEP-130130}}

\title{Dark Radiation Constraints \\
on Mixed Axion/Neutralino Dark Matter
}
\author{Kyu Jung Bae$^{a}$, Howard Baer$^{a}$ and Andre Lessa$^{b}$\\
$^a$Dept.\ of Physics and Astronomy, University of Oklahoma, Norman, OK 73019, USA\\
$^b$ Instituto de F\'isica, Universidade de S\~ao Paulo, S\~ao Paulo - SP, Brazil\\
E-mail: \email{bae@nhn.ou.edu}, \email{baer@nhn.ou.edu}, \email{lessa@fma.if.usp.br}, 
}

\abstract{
Recent analyses of CMB data combined with the measurement of BAO and $H_0$
show that dark radiation-- parametrized by the apparent number of 
additional neutrinos $\Delta N_{eff}$ contributing to the cosmic expansion--
is bounded from above by about $\Delta N_{eff}\alt 1.6$ at 95\% CL. 
We consider the mixed axion/neutralino cold dark matter scenario
which arises in $R$-parity conserving supersymmetric (SUSY) models 
wherein the strong CP problem is solved by hadronic axions 
with a concommitant axion($a$)/saxion($s$)/axino($\ta$) supermultiplet.
Our new results include improved calculations of thermal axion and saxion production 
and include effects of saxion decay to axinos and axions. 
We show that the above bound on $\Delta N_{eff}$ is easily satisfied if saxions are mainly
thermally produced and $m_{LSP} < m_{\ta} \lesssim m_s$. However, if the dominant
mechanism of saxion production is through coherent oscillations, the CMB data
provides a strong bound on saxion production followed by saxion decays to axions.
Furthermore we show that scenarios with mixed neutralino/axion dark matter are highly
constrained by combined CMB, BBN and Xe-100 constraints. In particular,
supersymmetric models with a standard overabundance of neutralino dark matter
are excluded {\it for all values of the Peccei-Quinn breaking scale}.
Next generation WIMP direct detection experiments may be able to discover or exclude mixed
axion-neutralino CDM scenarios where $s\to aa$ is the dominant saxion decay mode.
}
\keywords{Supersymmetry Phenomenology, Supersymmetric Standard Model, Dark Matter,
Axions}

\begin{document}

\section{Introduction}
\label{sec:intro}

The recent discovery of a Higgs-like boson with mass $m_h\simeq 125$ GeV at the CERN LHC
is an outstanding accomplishment~\cite{atlas_h,cms_h}, 
and seemingly provides the last of the matter states
encompassed within the Standard Model of particle physics. However, if indeed the new boson turns out
to be spin-0, then it raises a conundrum: how is it that fundamental scalar particles
can exist at or around the weak scale when their masses are quadratically divergent? 
The well-known solution is to extend the model to include {\it weak scale supersymmetry}, 
which reduces quadratic divergences to merely logarithmic while predicting the existence of a
panoply of new matter states: the so-called superpartners~\cite{wss}. Indeed, the new resonance has
its mass sitting squarely within the narrow window $m_h\sim 115-135$ GeV which is predicted 
by the minimal supersymmetric Standard Model, or MSSM~\cite{carena}.

One of the virtues of the MSSM is that it includes several candidates for particle dark matter.
In $R$-parity-conserving theories, it is common to conjecture the lightest neutralino
$\tz_1$ as CDM. Its relic abundance follows from thermal freeze-out in the early universe
and can be consistent with the observed dark matter abundance, although this imposes severe constraints on the MSSM
parameter space~\cite{susywimp}.

A different problem arises from QCD in that the chiral symmetry 
$U(2)_L\times U(2)_R\sim U(2)_V\times U(2)_A$ associated with two flavors of 
light quarks gives rise to $U(2)_V\to SU(2)_I\times U(1)_B$ (isospin and baryon number symmetry) 
plus a $U(2)_A$ whose breaking is expected to yield four light pions/pseudo-Goldstone bosons.
Weinberg conjectured the $U(1)_A$ symmetry is violated by quantum effects and indeed 'tHooft
showed this was so, explaining why $m_\eta\gg m_\pi$.
A consequence of 't Hooft's solution to the $U(1)_A$ problem is that the QCD Lagrangian
should contain the $CP$-violating term
\be
{\cal L}_{CPV}\ni \theta\frac{g_s^2}{32\pi}F_A^{\mu\nu}\tilde{F}_{A\mu\nu}
\ee
where $g_s$ is the QCD coupling, $F_{A\mu\nu}$ is the gluon field strength tensor ($\tilde F$ is its dual)
and $\theta$ is an arbitrary parameter. Measurements of the neutron EDM imply $\theta\alt 10^{-10}$
giving rise to the strong $CP$ problem: why is CP violation in the strong sector so small~\cite{review}?
After 35 years, still the most compelling solution is to invoke an additional Peccei-Quinn 
symmetry $U(1)_{PQ}$ whose breaking gives rise to the axion field $a$~\cite{pqww,ksvz,dfsz}: 
in this case, the offending ${\cal L}_{CPV}$ term dynamically relaxes to unobservable levels.
Phenomenology dictates the PQ symmetry is broken at a scale $f_a\sim 10^9-10^{16}$ GeV, where
the lower limit comes from astrophysical constraints~\cite{axreview} while the upper limit arises from
theoretical prejudice~\cite{witten}. The axion mass is given by
\be
m_a\simeq 6\ {\rm \mu eV}\left(\frac{10^{12}\ {\rm GeV}}{f_a}\right) .
\ee

Although its mass is predicted to be slight, the axion is still an excellent CDM candidate. It can be 
produced at temperatures around $1$ GeV via coherent oscillations with a relic abundance given by~\cite{Oh2axion}
\be
\Omega_a^{std} h^2\simeq 0.23 f(\theta_i)\theta_i^2 
\left(\frac{f_a}{10^{12}\ {\rm GeV}}\right)^{7/6}
\label{eq:Oh2axionstd}
\ee
where the misalignment angle $0< \theta_i<\pi$ and $f(\theta_i)$ is the anharmonicity
factor. Visinelli and Gondolo~\cite{vg1} parametrize the latter as
$f(\theta_i)=\left[\ln\left(\frac{e}{1-\theta_i^2/\pi^2}\right)\right]^{7/6}$.
The uncertainty in $\Omega_a^{std} h^2$ from vacuum mis-alignment is estimated 
as plus-or-minus a factor of three. 

In moving towards realistic models, it should be fruitful to incorporate
simultaneously solutions to both the gauge hierarchy problem and the strong $CP$ problem~\cite{susyaxion}.
In supersymmetric axion models, labeled here as the Peccei-Quinn-augmented MSSM or PQMSSM, 
the axion necessarily occurs as but one element of an axion supermultiplet
\be
\hat{a}=\frac{s+ia}{\sqrt{2}}+ i\sqrt{2}\bar{\theta}\ta_L +i\bar{\theta}\theta_L{\cal F}_a ,
\ee
in 4-component spinor notation~\cite{wss}. Here, we also introduce the 
$R$-parity-even scalar {\it saxion} field $s$ and the $R$-parity-odd spin-$1/2$ {\it axino}
field $\ta$. In gravity-mediation (as assumed in this paper) the saxion is
 expected to gain mass $m_s\sim m_{3/2}$, where $m_{3/2}$ is the
gravitino mass. More generally, the axino mass can range from keV to $m_{3/2}$~\cite{cl,ckkr,kim}.
Here we always assume $m_{\ta} \sim m_{3/2}$\cite{hall}, so the lightest suspersymmetric particle (LSP) is the neutralino.
The axion, saxion and axino couplings to matter are all suppressed by the PQ breaking scale 
$f_a$.
In the PQMSSM with a neutralino LSP ($m_{\ta} \sim m_{3/2}\sim TeV$), one
expects dark matter to be comprised of an axion-neutralino admixture~\cite{ckls,blrs,bls}.
In spite of their suppressed couplings to matter, both the axino and saxion can play surprising
roles in dark matter production rates in the early universe.

In a previous work, Choi {\it et al.} showed the effects of axino production
on the relic neutralino abundance~\cite{ckls}. Thermal production of axinos in the early universe
followed by their decays at temperatures
\be
T_D^{\ta} = \sqrt{\Gamma_{\ta}M_P}/(\pi^2g_*(T_D)/90)^{1/4}
\ee
(where $M_P = 2.4\times 10^{18}$ GeV is the reduced Planck mass and $\Gamma_{\ta}$ the axino decay width)
feeds additional neutralinos into the cosmic plasma if $T_D^{\ta} < T_{fr}$, 
where $T_{fr}$ is the neutralino freeze-out temperature. As a result, 
the neutralino abundance is augmented with respect
to the standard thermal (MSSM) abundance. However, if axinos are abundantly produce and come to
 dominate the universe at a temperature $T_e$, the
abundance of any relics present at the time of axino decay can be diluted by an entropy
injection factor $r =S_f/S_0\simeq T_e/T_D^{\ta}$. The value of $r$ can range from 1 (no dilution)
to values as high as $10^3-10^4$ or even higher~\cite{blrs} 
(see also Ref's~\cite{entropy}-\cite{bl}).
In Ref.~\cite{blrs}, this avenue was pushed much further, where analytic formulae were presented for both neutralino 
and axion production in either radiation-, matter- or decay-dominated universes. 
The effect of saxion production and entropy injection from late-time saxion decays was also
considered. In Ref.~\cite{bl}, entropy injection from coherent oscillation (CO)- produced saxions 
with decays  $s\to gg$ was found to strongly dilute all relics at the time of saxion decay. 
This effect allowed much higher values of $f_a\sim 10^{13}-10^{15}$ GeV to be cosmologically allowed,
even for $\theta_i$ as large as $\sim 0.1$.

While the semi-analytic approach presented in~\cite{ckls,blrs} and~\cite{bl} 
is applicable in many cases, there also exist numerous cases where a fully numeric solution 
to the coupled Boltzmann equations is required. 
Such cases include the possibility of bino-like neutralinos where $\langle\sigma v\rangle$ is
variable with temperature $T$ instead of constant, and where multiple processes of
neutralino injection or entropy dilution are possible, such as simultaneously accounting
for axino, axion, neutralino and saxion production and possible decays.
In Ref.~\cite{bls}, a coupled Boltzmann calculation was presented which tracked the
abundances of neutralinos, axions, saxions, axinos and gravitinos.
The calculations were restricted to the case where $s\to gg$ and $\tg\tg$ were dominant. 
It was found that SUSY models with a standard thermal underabundance of neutralinos 
(case of higgsino-like or wino-like LSPs) could easily have the neutralino abundance enhanced, 
and in addition any remaining
underabundance could be filled by relic axions. 
In such models, the value of $f_a$ could be pushed up to $10^{13}-10^{15}$ GeV owing to the combined
effects of neutralino enhancement and dilution from entropy production, as well as dilution of relic axions.
In the same work, it was found to be difficult to suppress
the neutralino abundance with respect to the standard thermal 
abundance in the MSSM.
At high values of $f_a$ where entropy dilution from CO-produced saxions was large,
a high level of dilution was usually accompanied by a violation of BBN bounds 
on late-decaying relics~\cite{ellis,kohri,jedamzik}, since it was assumed that most of
the saxion energy goes into visible energy through the $s \to gg,\tg\tg$ decays.
However this picture can be significantly modified once we consider possible $s \to aa,\ \ta\ta$ decays
as described below.

The interactions of saxions with axions and axinos is given by\cite{gs}
\be
{\cal L}\ni \left(1+\frac{\sqrt{2}\xi }{v_{PQ}}s\right)\left[\frac{1}{2}(\partial_\mu s)^2+
\frac{1}{2}(\partial_\mu a)^2 +\frac{i}{2}\bar{\ta}\dsl \ta \right] +\cdots
\label{eq:Ls}
\ee
with $\xi =\sum_iq_i^3v_i^2/v_{PQ}^2$ and where $q_i$ is the PQ charge of various PQ multiplets, 
$v_i$ are their vevs after PQ symmetry breaking and $v_{PQ}=\sqrt{\sum_i v_i^2q_i^2} = f_a/\sqrt{2}$.
In some simple models, $\xi$ can be small or even zero, while in others it can be as high as unity~\cite{cl}.
In the case where $\xi$ is non-zero, there is the possibility of additional decay modes of 
saxions which can influence the dark matter production rate. In particular, for $\xi \agt 0.05$,
the decay $s\to aa$ or $s\to \ta\ta$  can become relevant. The latter decay can feed additional
LSPs into the plasma, while the former decay gives rise to a population of relativistic axions
which forms the so-called {\it dark radiation}. 

Indeed, up until recently, 
cosmological data seemed to favor the existence of dark radiation,
not predicted by the Standard Model. The population of weakly interacting
relativistic degrees of freedom is parametrized by the number of effective neutrinos
($N_{eff}$), which is $\sim 3$ in the Standard Model, corresponding to the three neutrino flavors.
Previous data from WMAP7, the South Pole Telescope (SPT) and the Atacama Cosmology Telescope (ACT) suggested $N_{eff} \simeq 3.5-4.5$~\cite{wmap}, indicating a source
of dark radiation beyond the SM.
A variety of papers have recently explored this possibility~\cite{vb,ichikawa,hasen,hooper,gs,mel,conlon}.
More recently, the ACT~\cite{act} has released additional data, which reduced the $N_{eff}$ value combined with the measurement of baryon acoustic oscillations (BAO) and the Hubble constant to:
\begin{equation}
N_{eff}=3.50\pm0.42\qquad \mbox{(WMAP7+ACT+BAO+$H_0$)}.
\end{equation}
On the other hand, recent SPT~\cite{spt} and WMAP9~\cite{wmap9} analyses reported rather higher values,
\begin{eqnarray}
N_{eff}&=3.71\pm0.35\qquad &\mbox{(WMAP7+SPT+BAO+$H_0$)},\\
N_{eff}&=3.84\pm0.40\qquad &\mbox{(WMAP9+eCMB+BAO+$H_0$)}.
\end{eqnarray}
From the above numbers it is clear the tension between the latest ACT and SPT/WMAP9 values for $N_{eff}$. While the ACT result has only $1.1\sigma$-level deviation from the standard value, $N_{eff}=3.04$, the SPT and WMAP9 results show almost a $2\sigma$-level deviation\footnote{It is worth pointing out that in the WMAP9 analysis `eCMB' denotes the extended CMB, which uses the old data sets of SPT (2011) and ACT (2011). 
Also, each $N_{eff}$ value is obtained from different data sets for BAO and $H_0$.
Hence it is hard to determine the most updated result.
Meanwhile, in Ref.~\cite{divalentino}, an independent analysis was made, which consistently combines the most recent data sets from ACT and SPT with WMAP9 data.
The results obtained in this case for ACT+WMAP9+BAO+$H_0$ and SPT+WMAP9+BAO+$H_0$ are consistent with the latest values reported by ACT~\cite{act} and SPT~\cite{spt}.
}.
Due to the tension between the different analyses and the fact that all current results are compatible with the SM model value within $2\sigma$,
 it is hard to consider these results as a strong evidence for dark radiation.
Instead, we choose to use the current results as upper bounds on $N_{eff}$, which can be predicted by various models.
For the purpose of this paper, we will invoke a conservative constraint of
\be
\Delta N_{eff} \equiv N_{eff} - N_{eff}^{SM} < 1.6 \;.
\ee
Higher values of $\Delta N_{eff}$ are excluded at 95\% CL (or more) by any of the current CMB analysis discussed above\footnote{From here on we use `CMB' to encompass any of the current ACT, SPT or WMAP9 data analysis.}.


In this paper, we discuss the impact of the CMB constraint on the PQMSSM parameter space, 
assuming $m_{\ta} \sim m_{3/2}\sim$ TeV, so the neutralino is the LSP. 
In order to properly compute $\Delta N_{eff}$ and the cold axion and neutralino relic abundances, 
we numerically solve the coupled Boltzmann equations. 
Here we make several improvements with respect to the results in Ref.~\cite{bls}:
\bi
\item we include saxion decays to $aa$ and $\ta\ta$ final states,
\item we update thermal saxion production rates as recently calculated by Graf and Steffen~\cite{gs} and
\item we improve the system of Boltzmann equations to properly compute the amount of dark radiation.
\ei
While the decay $s\to aa$ may lead to a relativistic component of dark matter which is
strongly constrained by $\Delta N_{eff}<1.6$, it may also 
diminish the saxion entropy dilution effect which is large when $s\to gg$ dominates instead. Although most of our results 
are weakly dependent on the specific MSSM spectrum, for definiteness
we will apply our results to two benchmark SUSY models inspired by recent LHC results: one of
which contains a standard thermal overabundance (SOA) of bino-like neutralinos while the other
contains a standard underabundance (SUA) of higgsino-like neutralinos. 
As shown in Sec.~\ref{sec:num}, if $s\to aa$ decays dominate, the SOA case is typically excluded {\it for all $f_a$ values}
by combined BBN, dark radiation and dark matter constraints. For the SUA case, the Xe-100
direct WIMP search bound~\cite{xe100} also becomes relevant. For SUA,  some 
parameter points evade all constraints at low $f_a\sim 10^9-10^{11}$ GeV and also at high $f_a\sim 10^{15}-10^{16}$ GeV, 
but otherwise the bulk of parameter space points are also excluded by the combined constraints.
Next generation WIMP direct detection experiments may be able to discover or exclude mixed
axion-neutralino CDM scenarios with a SUA of neutralinos, if $s\to aa$ is the dominant saxion decay mode.

In the next section we define the two SUSY benchmark points used in our analysis. Then, in Sec.~\ref{sec:decay},
we discuss the main saxion and axino decay modes relevant for our results. Sec.~\ref{sec:boltz} introduces some
basic formalism and notation for our subsequent discussion and the calculation of $\Delta N_{eff}$ and the BBN constraints.
Analytical expressions for the dark radiation energy density in the PQMSSM are derived in Sec.~\ref{sec:darkrad}, while more general
results using numerical solutions are presented in Sec.~\ref{sec:num}. Finally, we present our conclusions in Sec.~\ref{sec:conclude}.
In an Appendix, we provide a detailed description of the formulae for the coupled Boltzmann equations 
used to compute our numerical solutions.

\section{Two SUSY benchmarks}
\label{sec:BMs}

In this Section, we present two SUSY model benchmark points which are useful
for illustrating the effects of dark radiation: one (labeled as SOA) has a standard thermal
{\it overabundance} of neutralinos, while the other (labeleld as SUA) has a standard thermal
{\it underabundance}.

For the SOA case, we adopt the mSUGRA/CMSSM model
with parameters
\be
 \mbox{$(m_0,\ m_{1/2},\ A_0,\ \tan\beta ,\ sign(\mu ))$ = $(3500\ {\rm GeV},\ 500\ {\rm GeV},\ -7000\ {\rm GeV},\ 10,\ +)$}
\ee
We generate the SUSY model spectra with Isajet 7.83~\cite{isajet}.
As shown in Table~\ref{tab:bm}, the SOA point  has
$m_{\tg}=1.3$ TeV and $m_{\tq}\simeq 3.6$ TeV, so it is beyond current
LHC sparticle search constraints. It is also consistent with the LHC Higgs discovery since $m_h=125$ GeV.
The lightest neutralino is mainly bino-like with $m_{\tz_1}=224.1$ GeV, and the
standard neutralino thermal abundance from IsaReD~\cite{isared} is found to be
$\Omega_{\tz_1}^{MSSM}h^2=6.8$, a factor of $\sim 60$ above the measured value~\cite{wmap9}.
Some relevant parameters, masses and direct detection cross sections 
are listed in Table~\ref{tab:bm}. Due to its heavy 3rd generation squark masses and large $\mu$ 
parameter, this point has very high electroweak finetuning~\cite{sug_ft}.

The second point listed as SUA comes from {\it radiative natural SUSY}~\cite{rns} 
with parameters from the 2-parameter non-universal Higgs model
\be
\mbox{$(m_0,\ m_{1/2},\ A_0,\ \tan\beta )$ = $(7025\ {\rm GeV},\ 568.3\ {\rm GeV},\ -11426.6\ {\rm GeV},\ 8.55)$}
\ee 
with input parameters $(\mu,\ m_A)=(150,\ 1000)$ GeV.
With $m_{\tg}=1.56$ TeV and $m_{\tq}\simeq 7$ TeV, it is also safe from LHC searches.
It has $m_h=125$ GeV and a higgsino-like neutralino with mass $m_{\tz_1}=135.4$ GeV 
and standard thermal abundance $\Omega_{\tz_1}^{MSSM}h^2=0.01$, low by about $\sim 10$
from the measured dark matter density. It has very low electroweak finetuning.

%
\begin{table}\centering
\begin{tabular}{lcc}
\hline
 & SOA (mSUGRA)  & SUA (RNS2)  \\
\hline
$m_0$ & 3500 & 7025 \\
$m_{1/2}$  & 500 & 568.3 \\
$A_0$ & -7000 & -11426.6 \\
$\tan\beta$  & 10 & 8.55 \\
$\mu $ & 2598.1 & 150 \\
$m_A$ & 4284.2 & 1000 \\
$m_h$ & 125 & 125.0 \\
$m_{\tg}$ & 1312 & 1562 \\
$m_{\tu}$ & 3612 & 7021 \\
$m_{\tst_1}$ & 669 & 1860 \\
$m_{\tz_1}$ & 224.1 & 135.4 \\
\hline
$\Omega^{std}_{\tz_1} h^2$ & 6.8 & 0.01 \\
$\sigma^{SI}(\tz_1 p)$ pb & $1.6\times 10^{-12}$ & $1.7\times 10^{-8}$ \\
\hline
\end{tabular}
\caption{Masses and parameters in~GeV units for two benchmark points
computed with \Isajet\,7.83 and using $m_t=173.2$ GeV.
}
\label{tab:bm}
\end{table}

\section{Saxion and Axino decays}
\label{sec:decay}

The partial widths for $s\to gg$ and $s\to\tg\tg$ have been discussed in several papers 
(see for instance Ref.~\cite{AY}). The decay $s\to\gamma\gamma$ is also
possible but is always subdominant and not of consequence here. Instead, we focus on the possibility of
$s\to aa$. Using the Lagrangian in Eq.~(\ref{eq:Ls}), we find
\be
\Gamma_{s\to aa}=\frac{\xi^2m_s^3}{64\pi v_{PQ}^2} = \frac{\xi^2m_s^3}{32\pi f_{a}^2} \label{eq:stoaa}
\ee
in accord with~\cite{gs}. Since it will be relevant for our subsequent discussion, we also list the
$s \to gg$ decay width:
\be
\Gamma_{s\to gg} = \frac{\alpha_s^2m_s^3}{32\pi^3 f_{a}^2} .
\label{eq:stogg}
\ee

In addition, we can consider the saxion-axino-axino interaction term
\be
{\cal L}\ni \frac{i\xi }{\sqrt{2}v_{PQ}}s\bar{\ta}\dsl\ta.
\ee
From this interaction Lagrangian, we obtain 
\be
\Gamma_{s\to\ta\ta}=\frac{\xi^2}{4\pi}\frac{m_{\ta}^2m_s}{f_{a}^2}\left(1-4\frac{m_{\ta}^2}{m_s^2}\right)^{3/2} .
\ee

It is worth noting that the $s\ta\ta$ coupling constant is not necessarily the 
same as the $saa$  one. 
Indeed, the $s\ta\ta$ coupling can be generated by both SUSY breaking and PQ symmetry breaking so that it is 
highly model-dependent.
The $s\ta\ta$ coupling can be much smaller than that of $saa$~\cite{choi}.
In this work, for simplicity, we assume a common coupling constant $\xi$, 
since it is sufficient to illuminate the effect of saxion decay to axino pairs. 
This assumption allows us to cover both cases where saxion decays to axion and to axino pairs.

\begin{figure}[t]
\begin{center}
\includegraphics[width=14cm]{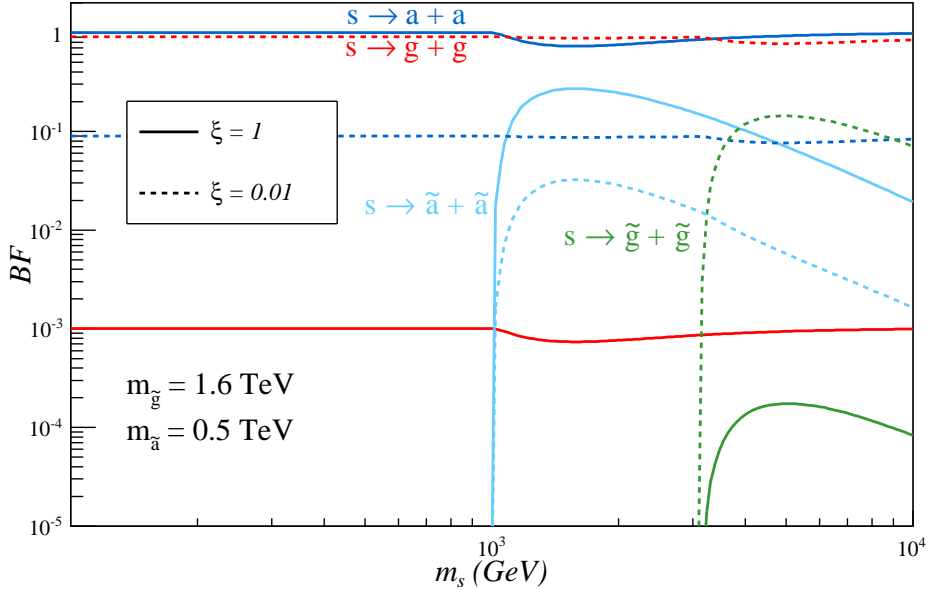}
\caption{Saxion branching fractions $s\to gg$ (red), $s\to\tg\tg$ (green), 
$s\to aa$ (dark blue) and $s\to\ta\ta$ (light blue) versus $m_s$ for $m_{\ta}=0.5$ TeV
and $m_{\tg}=1.6$ TeV. We take $\xi =0.01$ (dashed) and 1 (solid).
}
\label{fig:sBF}
\end{center}
\end{figure}

To show the relative importance of these decay modes, we show in Fig.~\ref{fig:sBF}
the saxion branching fractions vs. $m_s$ for the case where $m_{\ta}=0.5$ TeV and
$m_{\tg}=1.6$ TeV. We take $\xi =0.01$ (dashed curves) and $\xi =1$ (solid curves).
In the case of $\xi=0.01$, we see that $s\to gg$ dominates the saxion branching fraction
for all $m_s$ values. The coupled Boltzmann calculation of mixed $a\tz_1$ dark matter
presented in Ref.~\cite{bls} pertains to this case. If instead $\xi\sim 1$, then the 
situation changes radically, and we see that $s\to aa$ dominates for all values
of $m_s$. This decay mode, as mentioned earlier, may lead to substantial 
dark radiation and contribute to $N_{eff}$. In addition,  once $m_s>2m_{\ta}$ 
then $s\to\ta\ta$ turns on and rapidly becomes comparable to the $s\to aa$ branching fraction.
In this case, saxion decay followed by axino cascade decays may feed extra neutralinos
into the plasma, thus bolstering the neutralino abundance.
Since $\Gamma (s\to aa )\sim m_s^3$ while $\Gamma (s\to\ta\ta )\sim m_s$, then as $m_s$ increases
well past $2m_{\ta}$, the $s\to aa$ mode more completely dominates the saxion branching fraction.
%

Since here we assume $m_{\ta} > m_{\tz_1}$, the axino is unstable and decays to
$\tz_i + Z/\gamma$. Other decay channels to charginos and gluinos may also
be present if they are kinematically allowed. The specific expressions for each of
these decay modes have been discussed in detail in Ref.~\cite{blrs}. Although
all of these are included in our results, for most of our discussion the
only relevant decay mode is $\ta \to g + \tg$ with the decay width given by:
\be
\Gamma_{\ta \to \tg g} = \frac{\alpha_s^2}{16 \pi^3 f_a^2} m_{\ta}^3 \left(1 -\frac{m_{\tg}^2}{m_{\ta}^2}\right)^3 .
\ee
The above decay is the dominant one as long as it is kinematically allowed.

\section{Boltzmann equations for energy and number densities}
\label{sec:boltz}

Here we present a very brief description of our procedure to numerically calculate the
relic abundance of mixed $a\tz_1$ CDM in the PQMSSM and the amount of dark radiation ($\Delta N_{eff}$). A more detailed discussion
is left to Appendix~\ref{sec:appendix}.

\subsection{Boltzmann equations}

Under the assumptions described in Appendix~\ref{sec:appendix}, the general Boltzmann equations~\cite{kt} for the number density $n_i$ and energy density $\rho_i$
 of a particle species $i$ can be written as:
\bea
\Drv{n_i}{t} + 3 H n_i & = & - \Gamma_i m_i \frac{n_i^2}{\rho_i} + [(\bar{n}_{i}(T))^2 - n_{i}^2] \sigv_i + \sum_{j} BR(j \to i) \Gamma_j m_j \frac{n_j^2}{\rho_j} \label{eq:nieq} \\
\Drv{\rho_i}{t} + 3H (\rho_i + P_i) & = & \left( \bar{n}_i^2 - n_i^2 \right) \langle \sigma v \rangle \frac{\rho_i}{n_i}  -\Gamma_i m_i n_i + \sum_{j \neq i} \
BR(j \to i) \Gamma_j \frac{m_j}{2} n_j \label{eq:rieq}
\eea
where $P_i$ is the pressure density, $\Gamma_i$ the decay width, $\gamma_i$ is the relativistic dilation factor,
$\langle\sigma v\rangle$ is the  (temperature dependent) thermally averaged annihilation 
cross section times velocity for the particle species $i$, $\bar{n}_i$ is its equilibrium number density and
$BR(j \to i)$ is the branching fraction for particle $j$ to decay to particle 
$i$.\footnote{In this paper, particle species $i$ denotes
1. neutralinos $\tz_1$, 2. thermally produced (TP) axinos $\ta$, 3. and 
4. coherently produced (CO)- and TP-produced saxions $s(x)$, 
5. and 6. CO- and TP/decay-produced axions $a$, 
7. TP gravitinos $\tG$ and 8. radiation. 
We allow for axino decay to $g\tg$, $\gamma\tz_i$ and $Z\tz_i$ states ($i=1-4$), and
saxion decay to $gg$, $\tg\tg$, $aa$ and $\ta\ta$. 
Additional model-dependent saxion decays {\it e.g.} to $hh$ are
possible and would modify our results. 
We assume $\tG$ decay to all particle-sparticle pairs, and include 3-body gravitino 
modes as well~\cite{moroi_grav}.}
As discussed in Appendix~\ref{sec:appendix},
the above equation is also valid for coherent oscillating fields once we take $BR(j \to i)=0$ and
$\sigv_i = 0$.

It is also convenient to write an equation for the evolution of entropy ($S$):
\bea
\dot{S} & = &\left(\frac{2 \pi^2}{45} g_*(T) \frac{1}{S}\right)^{1/3} R^4 \sum_{i} R(i\to X) \frac{1}{\gamma_i} \Gamma_i \rho_{i} \nonumber \\
{\rm or}\ \ \  \dot{S} & = &\frac{R^3}{T} \sum_{i} R(i\to X) \Gamma_i m_i n_{i} 
\label{eq:Seq}
\eea
where $R(i\to X)$ is the fraction of energy injected in the thermal bath from $i$ decays.
 
We supplement the above with Friedmann's equation:
\be
H = \frac{1}{R} \frac{dR}{dt} = \sqrt{\frac{\rho_T}{3 M_P^2}} \; , \mbox{ with } \rho_T \equiv \sum_{i} \rho_i + \frac{\pi^2}{30} g_*(T) T^4 \; ,
\label{eq:H}
\ee
where $M_P$ is the reduced Planck mass.
The set of coupled differential equations, Eqs.~(\ref{eq:nieq}), (\ref{eq:rieq}), (\ref{eq:Seq}) and (\ref{eq:H})
can be solved as a function of time.
At late times ($T \ll 1$ MeV), when unstable particles have decayed (except for the axion) and
the neutralino has decoupled, the CDM relic abundance is given by:
\be
\Omega_{CDM} = \frac{\rho_{\tz_1}(T)}{\rho_c} \frac{s(T_0)}{s(T)} + \frac{\rho_{a}^{CO}(T)}{\rho_c} \frac{s(T_0)}{s(T)} 
\ee
where $\rho_{a}^{CO}$ is the energy density of cold (coherent oscillating) axions,
 $T_0$ is today's temperature and $\rho_c$ is the critical density. Furthermore, since dark radiation is composed
of hot axions (both thermally and non-thermally produced), we have:
\be
\Delta N_{eff}  = \frac{\rho_a(T)}{\rho_{\nu}}= \frac{\rho_{a}(T)}{T^4} \frac{120}{7 \pi^2} \left(\frac{11}{4}\right)^{4/3} \label{eq:dneff0}
\ee
where we used $\rho_{\nu} = \frac{7}{8}\frac{\pi^2}{15}T_{\nu}^4$ and $T_{\nu} = \left(\frac{4}{11}\right)^{1/3} T$.

\subsection{Constraints from BBN and Dark Radiation} 
\label{ssec:bbn}

%
A critical constraint on unstable relics comes from maintaining the success of the standard picture of Big Bang nucleosynthesis. 
Constraints from BBN on late decaying neutral particles (labeled $X$) have been
calculated recently by several groups~\cite{ellis,kohri,jedamzik}.
We have constructed digitized fits to the constraints given in Ref.~\cite{jedamzik} and apply these to late decaying
gravitinos, axinos and saxions. Typically, unstable neutrals with decay temperature below
5~MeV (decaying during or after BBN) and/or large abundances will be more likely to destroy
the predicted light element abundances. We point out that BBN also constrains $N_{eff}$ since it affects
the time of neutron freeze-out and consequently the $He/H$ ratio. 
However this constraint is usually weaker than the CMB constraint we assume here ($\Delta N_{eff} < 1.6$).

\subsection{Sample calculation from benchmark SUA: radiative natural SUSY} 
\label{ssec:example}

As an example calculation, we adopt the benchmark point SUA from Sec.~\ref{sec:BMs} 
with a higgsino-like neutralino and a standard neutralino underabundance of neutralino
dark matter. 

\begin{figure}[t]
\begin{center}
\includegraphics[width=14cm]{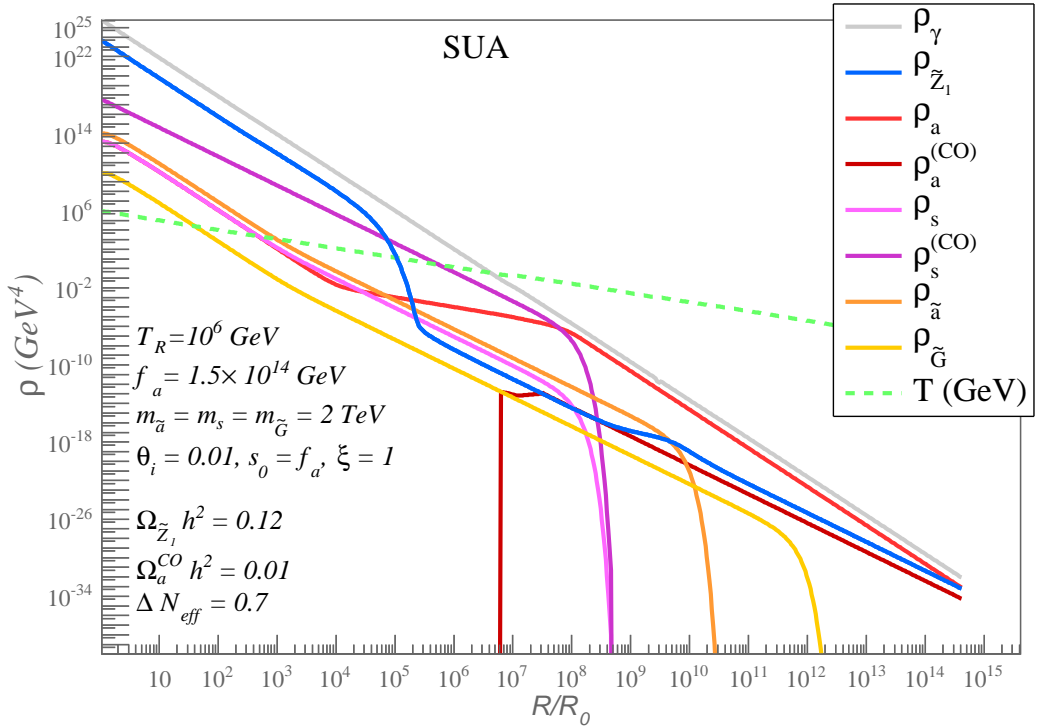}
\caption{Evolution of various energy densities versus scale 
parameter $R/R_0$ for the SUA benchmark.
}
\label{fig:RNS2}
\end{center}
\end{figure}

Working in the hadronic axion PQMSSM framework, we assume $T_R=10^6$ GeV with PQ parameters 
$m_{\ta}=m_s=m_{\tG}=2$ TeV, $\theta_i=0.01$ and $\fa =1.5\times 10^{14}$ GeV. 
We take $s_0 = f_a$, where $s_0$ is the initial field amplitude for
coherent oscillating saxions.
We also take $\xi =1$ so that $s\to aa$ is the dominant saxion decay mode.
The various energy densities $\rho_i$ are shown in Fig.~\ref{fig:RNS2}
for $i=\gamma$ (radiation), $\tz_1$ (neutralinos), $a$ (combined thermally- and decay-produced axions), 
$a^{CO}$ (coherent oscillating axions), $s$ (thermally produced saxions), $s^{CO}$ (coherent oscillating
saxions), $\ta$ (thermally produced axinos)
and $\tG$ (thermally produced gravitinos). The energy densities are plotted against the scale factor ratio $R/R_0$, 
where $R_0$ is the scale factor at $T=T_R$. We also plot the temperature $T$ (in GeV) of
radiation (green-dashed curve). 
%

We see that at all values of $R/R_0$ the universe in this case is radiation-dominated.
At $T \gg 1$ TeV, the TP axions, saxions and axinos all have similar abundances. At
these temperatures, the saxion coherent abundance is well above these components while the gravitino
thermal abundance is far below the other components. 
At low $R/R_0$, the neutralinos are in thermal equilibrium and their energy density lies
well above the abundances of members of the PQ multiplet. 
As the universe expands and cools, most components
are relativistic, and decrease with the same slope as radiation: $\rho_i\sim T^{-4}$. The exception 
is the CO-produced saxions, which behave as a non-relativistic fluid and fall-off as $\rho_{s}^{CO}\sim T^{-3}$.
At $R/R_0\sim 10^3$, the thermally-produced (TP) axinos, saxions and gravitinos become non-relativistic, 
so now $\rho_{\ta,s,\tG}^{TP}\sim T^{-3}$. For slightly lower temperatures with
$R/R_0\sim 10^5$, neutralinos begin to freeze-out, and their abundance falls steeply.
At $R/R_0\sim 10^{9}$, axinos begin to decay and bolster the neutralino abundance.
More importantly, at $R/R_0\sim 10^4$, the decay-produced axion abundance begins to swell
due to saxion decay until by $R/R_0\sim 10^8$ the decay-produced axion energy density is
nearly equal to the radiation density.
Also, around $R/R_0\sim 10^{7}$ with $T\sim 1$ GeV, CO production of axions begins, and by
$R/R_0\sim 10^{8}$, with $T\ll \Lambda_{QCD}$, its energy density begins to fall as $T^{-3}$. 
For the case illustrated here, the final neutralino abundance turns out to be
$0.12$ while the final axion density turns out to be $0.01$, due to the small value of $\theta_i$ chosen. The dark radiation from 
decay produced axions turns out to give $\Delta N_{eff}=0.7$ and would be allowed by the CMB constraint assumed here.
Below we discuss how the dark radiation constrains the PQMSSM parameters in more general scenarios.

\section{Dark radiation in the PQMSSM}
\label{sec:darkrad}

As seen in Fig.~\ref{fig:sBF}-- unless $\xi \ll 1$-- saxions mainly decay to axions. 
Since $m_s \gg m_a$ and $m_a \lesssim $ meV, the (non-thermally produced) 
axions injected from saxion decay remain relativistic until very late 
and contribute to $N_{eff}$. Furthermore, axions can be thermally produced 
(in equilibrium or out of equilibrium) after reheating and will also contribute as an 
additional relativistic species.
Below we discuss under which conditions TP or non-thermally produced (decay-produced) axions 
lead to an excess in $N_{eff}$ since such scenarios are severely constrained by the new CMB results.
First we present approximate analytical expressions for $\Delta N_{eff}$ in order to qualitatively 
discuss its dependence on the PQ parameters. 
We then use the full set of Boltzmann equations to numerically compute $\Delta N_{eff}$ and 
obtain our final results.

When computing the abundance of relativistic axions at late times ($T \ll 1$ MeV), two competing effects
must be taken into consideration: axion injection from saxion decays and entropy injection from both 
saxion ($s \to gg,\tg\tg$) and axino decays. 
While the former enhances the amount of dark radiation, the latter dilutes it. Thus, in the case
of entropy dilution, Eq.~(\ref{eq:dneff0}) becomes:
\be
\Delta N_{eff} = \frac{1}{r} \frac{\rho_{a}(T)}{\rho_{\nu}(T)} = \frac{1}{r} \frac{\rho_{a}(T)}{T^4} \frac{120}{7 \pi^2} \left(\frac{11}{4}\right)^{4/3} \label{eq:dneff}
\ee
where $r$ is the entropy
dilution factor. The above expression must be computed at $T \sim$ eV, which are the temperatures to which the CMB is sensitive.

As mentioned before, axions  can be both thermally and non-thermally produced in the early universe, so $\rho_a$ 
receives contributions from thermal production of axions and thermal and coherent production of saxions, followed by $s \to aa$ decays.
Thus the axion energy density after saxion decays is:
\be
\rho_a(T) = \rho_a^{TP}(T) + BR(s\to aa) \left(\frac{g_{*S}(T)}{g_{*S}(T_D)}\right)^{4/3} \left(\frac{T}{T_D}\right)^4 \rho_s(T_D) \label{eq:rhoa}
\ee
where $T_D$ is the saxion decay temperature. In the above expression we do not include the possibility of entropy dilution of $\rho_a$, since
this is accounted for by the dilution factor $r$ in Eq.~(\ref{eq:dneff}). Using
\be
\rho_s(T_D) = m_s Y_s s(T_D) \mbox{ and } g_{*S}(T\sim eV) = 3.9 \, ,
\ee
where $s(T) = 2 \pi^2 g_{*S}(T) T^3/45$ is the entropy density and combining Eqs.~(\ref{eq:dneff}) and (\ref{eq:rhoa}), we obtain:
\be
\Delta N_{eff} \simeq \frac{1}{r}\frac{\rho_a^{TP}}{\rho_{\nu}} + 18.02 \frac{1}{r} BR(s\to aa) g_{*S}(T_D)^{-1/3} \frac{m_s Y_s}{T_D} \label{eq:dneff2}
\ee

The dilution of relics due to entropy injection of saxion and axino decays has been extensively discussed in the literature 
(see Refs.~\cite{blrs,bl} and references therein). In the general case both saxions and axinos can dominate the universe and inject
entropy at different times, which can lead to quite involved scenarios. These will be properly addressed once we present our numerical
results in Sec.~\ref{sec:num}. For our analytical results we assume that either saxion or axinos inject entropy (but not both), so we can approximate the entropy
dilution factor by:
\be
r = \max\left[1,\frac{4}{3} R(\ta \to X) \frac{m_{\ta} Y_{\ta}}{T_D^{\ta}} + \frac{4}{3} R(s \to X) \frac{m_{s} Y_{s}}{T_D}\right] \, , \label{eq:r}
\ee
and we will later assume that either the axino or the saxion term dominates. In the above expression 
$T_D^{\ta}$ and $Y_{\ta}$ are the axino decay temperature and yield and $R(i \to X)$ is the fraction of $\rho_i$ which goes
into visible energy (photons, leptons and jets). 
The above expression ensures that $r \geq 1$, so it is valid even when there
is no entropy dilution ($r=1$).
Below we review the analytical expressions necessary to compute $\rho_a^{TP}$, $Y_s$ and $r$ in order to obtain $\Delta N_{eff}$ using Eq.~(\ref{eq:dneff2}).

The thermal production of axions and saxions (in supersymmetric axion models) was calculated in 
Ref.~\cite{gs} and is given by (for relativistic axions):
\be
\frac{\rho_a^{TP}}{s} \simeq 5.8 \times 10^{-6}  g_s^6 \ln\left(\frac{1.01}{g_s}\right)\left(\frac{10^{12} \mbox{ GeV}}{f_a}\right)^2 \left(\frac{T_R}{10^8 \mbox{ GeV}}\right) g_{*S}^{1/3}(T) T \label{eq:tp}
\ee
where $s = 2 \pi^2 g_{*S}(T) T^3/45$ is the entropy density. 
Saxions are also thermally produced at the same rate~\cite{gs}, but since $m_s \gg m_a$, saxions become 
non-relativistic and decay in the early universe. 
Hence their energy density before decay is given by:
\be
Y_{s}^{TP} m_{s} = \frac{\rho_s^{TP}}{s} \simeq 1.33 \times 10^{-5}  g_s^6 \ln\left(\frac{1.01}{g_s}\right)\left(\frac{10^{12} \mbox{ GeV}}{f_a}\right)^2 \left(\frac{T_R}{10^8 \mbox{ GeV}}\right) m_s .
\label{eq:tp2}
\ee
In order to compute $r$, it is also necessary to know the thermal production of axinos, 
which has been computed in Ref.~\cite{strumia}:\footnote{We stress that the thermal production of 
saxions and axinos implemented here are calculated in different ways.
The saxion yield from Ref.~\cite{gs} is obtained from the Hard Thermal Loop (HTL) approximation method, 
while the axino yield from Ref.~\cite{strumia} is obtained from finite temperature field theory.
According to Ref.~\cite{strumia}, the HTL yield may be smaller than the finite-temperature calculation 
by factors as large as $\sim 4$ for low reheat temperatures ($T_R \sim 10^6$ GeV), when the strong coupling constant $g_s\sim 1$ and the HTL calculation is no longer valid.
Since a finite-temperature calculation for the axion/saxion yield is not presently available, we use the result of Ref.~\cite{gs}, even for low $T_R$ values.
}
\be
Y_{\ta} m_{\ta} = \frac{\rho_{\ta}^{TP}}{s} \simeq 0.9 \times 10^{-5}  g_s^6 \ln\left(\frac{3}{g_s}\right)\left(\frac{10^{12} \mbox{ GeV}}{f_a}\right)^2 \left(\frac{T_R}{10^8 \mbox{ GeV}}\right) m_{\ta} .
\label{eq:tp3}
\ee

We point out that Eqs.~(\ref{eq:tp})-(\ref{eq:tp3}) assume out of equilibrium production of 
axions, saxions and axinos, being valid only if $T_R$ is smaller than the decoupling temperature ($T_{dec}$). 
However, if $T_R > T_{dec}$, the axion and saxion number densities
are given by their thermal equilibrium values:
\be
\frac{\bar{\rho}_a}{s} \simeq 5.3 \times 10^{-4} g_{*S}(T)^{1/3} T \mbox{ , } \;\; \frac{\bar{\rho}_s}{s} \simeq 1.2 \times 10^{-3} m_s \;\; \mbox{ and } \;\;  \frac{\bar{\rho}_{\ta}}{s} \simeq 1.8 \times 10^{-3} m_{\ta}  \label{eq:tpeq}
\ee
From Eqs.~(\ref{eq:tp2}) and (\ref{eq:tpeq}), we can estimate the decoupling temperature:\footnote{Due to the different calculation methods
used to compute the non-thermal production of saxion/axions and axinos mentioned above,
axinos decouple at slightly smaller temperatures than axions and saxions, but for simplicity here we take a common decoupling temperature for axions, saxions and axinos.}
\be
T_{dec} \simeq 1.4 \times 10^{11} \mbox{ GeV} \left(\frac{f_a}{10^{12} \mbox{ GeV}}\right)^2 . 
\label{eq:tdec}
\ee

Besides being thermally produced, saxions can also be produced through coherent oscillations, 
resulting in the following energy density:
\be
Y_s^{CO} m_s = \frac{\rho_s^{CO}}{s} \simeq 1.9 \times 10^{-5} \mbox{ GeV} \frac{\min[T_R,T_s]}{10^8\ {\rm GeV}} \left(\frac{s_0}{10^{12}\ {\rm GeV}}\right)^2 \label{eq:co}
\ee
where $T_s$ is the temperature at which saxions start to oscillate, given by $3H(T_s) = m_s$. 
In the above expression, $s_0$ is the initial saxion field amplitude
and depends on the UV details of the model and the inflation dynamics. 
Natural scales for $s_0$ are usually taken to be of order the PQ breaking scale $f_a$ 
or the reduced Planck mass $M_{P}$~\cite{kkn,moroi}.

Finally, to compute $\Delta N_{eff}$ we also need the axino and saxion decay temperatures. Assuming that these fields decay in a radiation
dominated universe\footnote{It is also possible that saxions and/or axinos decay in an axino or saxion dominated universe,
which modify the values of $T_D$ and $T_D^{\ta}$. However, in this section
we neglect these effects. Once we present our numerical results in Sec.~\ref{sec:num}, these effects are automatically taken into account.}, we have:
\be
T_D^{\ta} = \sqrt{\Gamma_{\ta}M_P}/(\pi^2g_*(T_D^{\ta})/90)^{1/4} \;\; \mbox{ and } \;\; T_D = \sqrt{\Gamma_{s}M_P}/(\pi^2g_*(T_D)/90)^{1/4} \label{eq:tds}
\ee
where $\Gamma_{\ta}$ and $\Gamma_s$ are the axino and saxion widths, respectively.

Using the expressions presented above, we can compute the amount of dark radiation in PQMSSM models:
\be
\Delta N_{eff} \simeq \Delta N_{eff}^{TP}  + 18.02 \frac{1}{r} BR(s\to aa) g_{*S}(T_D)^{-1/3} \frac{m_s (Y_s^{CO} + Y_s^{TP})}{T_D} \label{eq:dneff2c}
\ee
where $\Delta N_{eff}^{TP} \equiv \frac{1}{r} \rho_a^{TP}/\rho_{\nu}$, $Y_s^{CO}$ and $Y_s^{TP}$ are the coherent oscillation and
thermal yields of saxions and 
\be
r = \max\left[1,\frac{4}{3} R(\ta \to X) \frac{m_{\ta} Y_{\ta}^{TP}}{T_D^{\ta}} + \frac{4}{3} R(s \to X) \frac{m_{s} (Y_{s}^{TP} + Y_s^{CO})}{T_D}\right] .
\ee
Before discussing the dark radiation constraints to the PQMSSM parameter space, we point out that the first term
in Eq.~(\ref{eq:dneff2c}) is always subdominant.
The maximum contribution of thermally produced axions happens
when these are produced in equilibrium ($\rho_a^{TP} = \bar{\rho}_a$) and there is no entropy dilution ($r=1$), which results in the following
upper bound:
\be
\Delta N_{eff}^{TP} \leq \frac{\bar{\rho}_{a}(T)}{T^4} \frac{120}{7 \pi^2} \left(\frac{11}{4}\right)^{4/3} \simeq 9.5 \times 10^{-3}
\ee
where we used Eq.~(\ref{eq:tpeq}) with $g_{*S}(T \sim eV) = 3.9$. 
As we can see, the contribution from TP axions is well below the current experimental sensitivity and can be safely neglected in our
subsequent discussion. Hence, in the results below, we only consider the saxion contribution to $\Delta N_{eff}$:
\be
\Delta N_{eff} \simeq 18.02 \frac{1}{r} BR(s\to aa) g_{*S}(T_D)^{-1/3} \frac{m_s (Y_s^{CO} + Y_s^{TP})}{T_D} .
\label{eq:dneff2b}
\ee

In most cases, $\Delta N_{eff}$ is dominated either by $Y_{s}^{CO}$ or $Y_{s}^{TP}$, so it is interesting to separately
discuss each of these scenarios. In Sec.~\ref{sec:tp}, we discuss the case where the main contribution to $\Delta N_{eff}$ comes
from thermal production of saxions ($Y_s \simeq Y_{s}^{TP}$), while in Sec.~\ref{sec:co} we discuss the case where $Y_s \simeq Y_{s}^{CO}$.
Then, in Sec.~\ref{sec:num}, we present our numerical results which do not assume the sudden decay approximation, properly take into
account the effects of axino and/or saxion dominated universes and cover all possible scenarios, 
including $Y_s^{CO} \simeq Y_s^{TP}$.

\subsection{$\Delta N_{eff}$ from Thermal Production}
\label{sec:tp}

Here we assume that saxions are mostly thermally produced, so $Y_s^{TP} \gg Y_s^{CO}$. This happens for large $T_R$ and/or small $s_0$.
In this case, Eq.~(\ref{eq:dneff2b}) becomes:
\be
\Delta N_{eff} \simeq 18.02 \frac{1}{r} BR(s\to aa) g_{*S}(T_D)^{-1/3} \frac{m_s Y_s^{TP}}{T_D} .
\label{eq:dnefftp}
\ee
We also assume that $\xi \gtrsim 0.05$, so saxion decays to gluons and gluinos are suppressed (see Fig.~\ref{fig:sBF}). 
In this case, entropy injection is dominated by axino decays, so that
\be
r \simeq \max\left[1,\frac{4}{3} \frac{m_{\ta} Y_{\ta}^{TP}}{T_D^{\ta}}\right]  ,
\label{eq:rtp}
\ee
where we have assumed $R(\ta \to X) \simeq 1$ which is usually a good approximation unless $m_{\ta} \simeq m_{\tz_1}$.

We will show below that thermal production of saxions usually gives $\Delta N_{eff} \ll 1$, so the CMB constraint
is easily satisfied in this case. In order to show this, it is sufficient to compute an upper bound for $\Delta N_{eff}$.
From Eqs.~(\ref{eq:dnefftp}) and (\ref{eq:rtp}) we have:
\be
\Delta N_{eff} \leq 18.02 BR(s\to aa) g_{*S}(T_D)^{-1/3} \frac{m_s Y_s^{TP}}{T_D}/(\frac{4}{3} \frac{m_{\ta} Y_{\ta}^{TP}}{T_D^{\ta}})
\ee
Note that the equality is satisfied in the case of entropy dilution ($r > 1$), while it overestimates $\Delta N_{eff}$ if $r=1$.
Now, from the expressions for the thermal production of saxions and axinos, 
Eqs.~(\ref{eq:tp2}), (\ref{eq:tp3}) and (\ref{eq:tpeq}) we have:
\be
Y_s^{TP}/Y_{\ta}^{TP} \leq \bar{Y}_{s}/\bar{Y}_{\ta} = \frac{2}{3}
\ee
where $\bar{Y}_{i}$ is the thermal equilibrium yield for $i$. Hence:
\bea
\Delta N_{eff} & \leq & 9.01  BR(s\to aa) g_{*S}(T_D)^{-1/3} \frac{m_s}{m_{\ta}} \frac{T_D^{\ta}}{T_D} \nonumber \\
  & = &  9.01  BR(s\to aa) g_{*S}(T_D)^{-1/12} g_{*S}(T_D^{\ta})^{-1/4} \frac{m_s}{m_{\ta}} \sqrt{\frac{\Gamma_{\ta}}{\Gamma_s}} \label{eq:dnefftp2}
\eea
where we have used Eq.~(\ref{eq:tds}) for the axino and saxion decay temperatures. 
In order to simplify even further the above expression,
we just need to compute $\Gamma_{\ta}/\Gamma_s$. 
From Fig.~\ref{fig:sBF} we see that the $s \to \tg \tg$ is always subdominant, while both $s \to aa$ and $s \to gg$ can be dominant, depending on the
value of $\xi$. Furthermore, unless $m_s \sim 2 m_{\ta}$, the decay to axinos can also be neglected, which we do here for simplicity.
Hence we can approximate the total saxion width by:
\be
\Gamma_s \simeq \frac{m_s^3}{32 \pi f_a^2} \left( \xi^2  + \frac{\alpha_s^2}{\pi^2} \right) \label{eq:gammas}
\ee
which gives the following branching ratio for the decay to axions:
\be
BR(s\to aa) \simeq \frac{\xi^2}{\xi^2  + \alpha_s^2/\pi^2} \label{eq:br}
\ee

As discussed in Sec.~\ref{sec:decay}, the axino decay width depends on the gaugino spectrum and the axino mass. 
Several axino decay modes have been computed in Ref.~\cite{blrs} where it has been shown that 
the axino decay width is dominated by $\ta \to g\tg$, if $m_{\ta} \gg m_{\tg}$. Thus, if we take 
the limit $m_{\tg} \to 0$, it is easy obtain an upper bound for $\Gamma_{\ta}$:
\be
\Gamma_{\ta} \leq \Gamma_{\ta \to g\tg}(m_{\tg} \to 0) = \frac{\alpha_s^2}{16 \pi^3 f_a^2} m_{\ta}^3 .
\ee
Finally, using the above results in Eq.~(\ref{eq:dnefftp2}), we have:
\be
\Delta N_{eff} \leq (0.09-0.19) \frac{\xi^2}{\left(\xi^2  + \alpha_s^2/\pi^2\right)^{3/2}} \sqrt{\frac{m_{\ta}}{m_s}} \label{eq:ntpmaxr}
\ee
where the above range corresponds to $g_{*S}(T_D), g_{*S}(T_D^{\ta}) = 10-100$.
In Eq.(\ref{eq:ntpmaxr}), the equality is satisfied only if $T_R > T_{dec}$, $r>1$ and $m_{\ta} \gg m_{\tg}$.
Otherwise, $\Delta N_{eff}$ is below the quoted value.
Nonetheless the above result illustrates the fact that for models with $\xi \simeq 1$ and $m_{\ta} < m_s$,
the CMB constraints are automatically satisfied. It is interesting to notice that, although  smaller values of $\xi$
suppress $BR(s\to aa)$, the amount of dark radiation actually increases, due to the
increase in the saxion lifetime, which compensates the decrease on the branching ratio.
We also point out that although the above bound is quite general and independent of all other PQ paramaters, 
it is only valid for {\it thermal production of saxions}.
As we will see in the next section, this result drastically changes
once we consider coherent production of saxions.

In order to verify the above results, we numerically compute $\Delta N_{eff}$ as a function of 
$f_a$ using the coupled Boltzmann equations for the PQ fields, but neglecting the contribution from
CO saxions.
We take $m_{s} = 1$ TeV, $T_R = 10^{10}$ GeV, $\xi = 1$ and $m_{\ta} = 3$ and 32 TeV.
In Fig.~\ref{fig:neff1}, the solid blue (red) line shows the full numerical solution for $m_{\ta} = 3$ (32) TeV, 
while the dashed lines show the respective upper bound
from Eq.~(\ref{eq:ntpmaxr}) using the appropriate values of $g_{*S}(T_D)$ and $g_{*S}(T_D^{\ta})$. 
The dashed gray line shows the amount of entropy dilution from axino decays for $m_{\ta} = 3$ TeV.
As we can see, $\Delta N_{eff}$ is always below the upper limit from Eq.~(\ref{eq:ntpmaxr}) even
when the full numerical solution is considered.
Furthermore, for most values of $f_a$ it is well below the bound 
since in these regions we have $r = 1$ and/or $T_R < T_{dec}$, 
where  Eq.~(\ref{eq:ntpmaxr}) is too conservative.

%
\FIGURE[t]{
\includegraphics[width=13cm,clip]{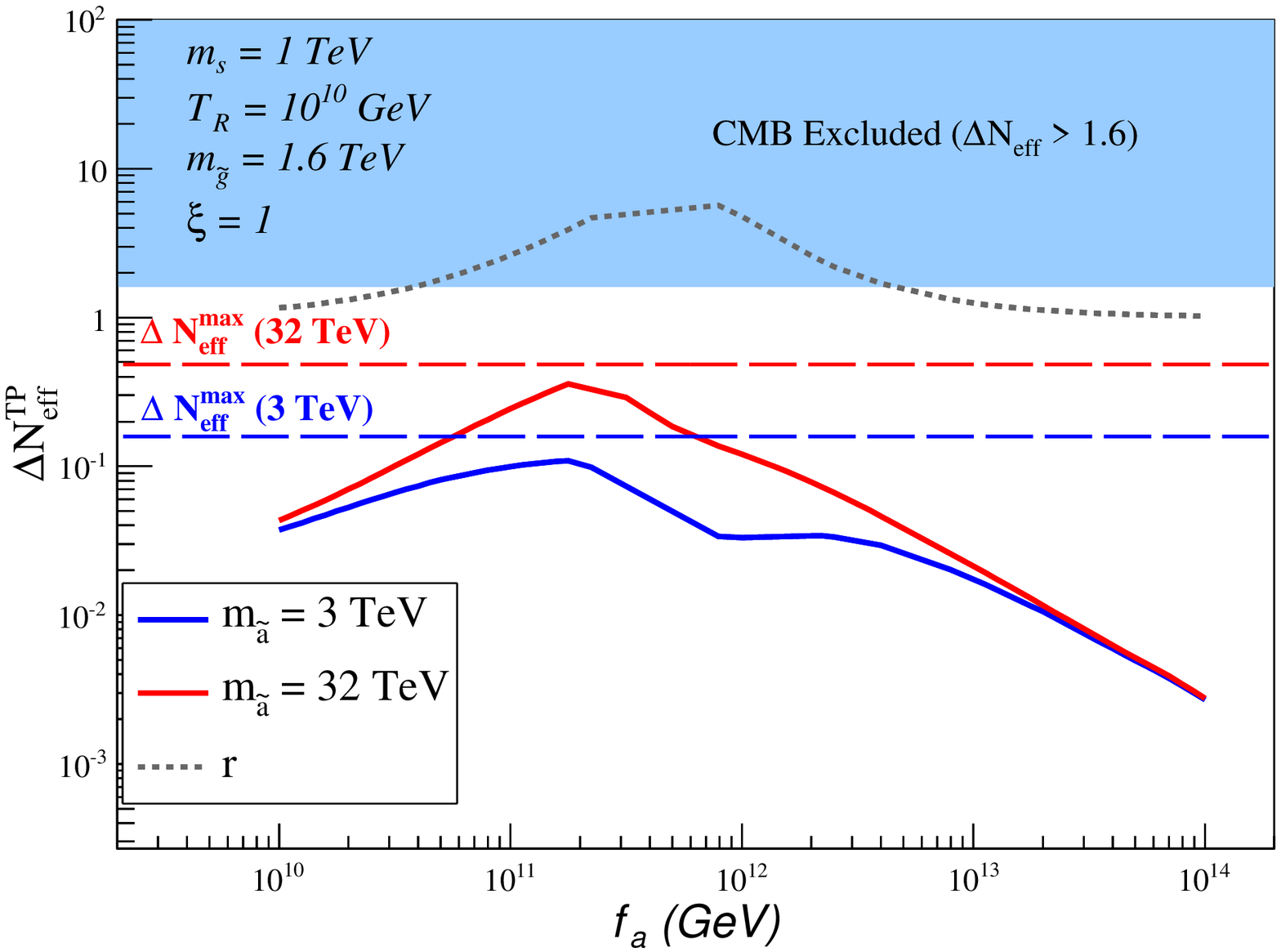}
\caption{$\Delta N_{eff}$ as a function of the PQ breaking scale $f_a$ for thermal production of saxions only. 
The fixed PQ parameters are $m_s = 1$ TeV, $T_R = 10^{10}$ GeV and $\xi =1$. 
The solid blue (red) line corresponds to the numerical solution for $m_{\ta} = 3$ (32) TeV.
The dashed lines correspond to the respective analytical upper bounds given by Eq.~(\ref{eq:ntpmaxr}). 
The entropy dilution factor ($r$) for $m_{\ta} = 3$ TeV is shown by the dotted gray line. 
The light blue region has $N_{eff} > 1.6$ and is excluded at 95\% C.L. by the CMB results.}
\label{fig:neff1}}

The results from Eq.~(\ref{eq:ntpmaxr}) and Fig.~\ref{fig:neff1} show that in order to violate the CMB constraint
on dark radiation ($\Delta N_{eff} < 1.6$),
the axino needs to be at least {\it two orders of magnitude} heavier than the saxion, if $\xi = 1$.
This is hard to achieve on most supersymmetric PQ models, where typically $m_{\ta} \leq m_{3/2} \sim m_{s}$~\cite{kim}.
For smaller $\xi$ values, $\Delta N_{eff}$ increases but it is still below the CMB bound as long as $m_{\ta}\lesssim m_s$.
To illustrate this we show in Fig.~\ref{fig:xibounds} the maximum allowed value of $m_{\ta}/m_s$ as a function of
$\xi$ according to the analytical result of Eq.~(\ref{eq:ntpmaxr}).
The region below the curve satisfies $\Delta N_{eff} < 1.6$, {\it irrespective of the other PQ parameter values},
as indicated by Eq.~(\ref{eq:ntpmaxr}). On the other hand the region above the curve can be either allowed or
excluded depending on the choice of PQ parameters. As shown in Fig.~\ref{fig:xibounds}, the CMB constraint
can be easily satisified for any value of $\xi$ if $m_{\ta} < 2m_s$. We also point out that if the current ACT~\cite{act} 
allowed interval for $N_{eff}$ ($\Delta N_{eff} < 1.3$) is assumed, it can still be satisfied for any value of $\xi$ as long as $m_{\ta} < 1.4m_s$.
Therefore, we conclude that the CMB constraint on $\Delta N_{eff}$ can be easily accomodated in the PQMSSM
if {\it saxions are mainly thermally produced}. In the following we discuss the case in which saxion production
is dominated by its coherent oscillation component.

\FIGURE[t]{
\includegraphics[width=11cm,clip]{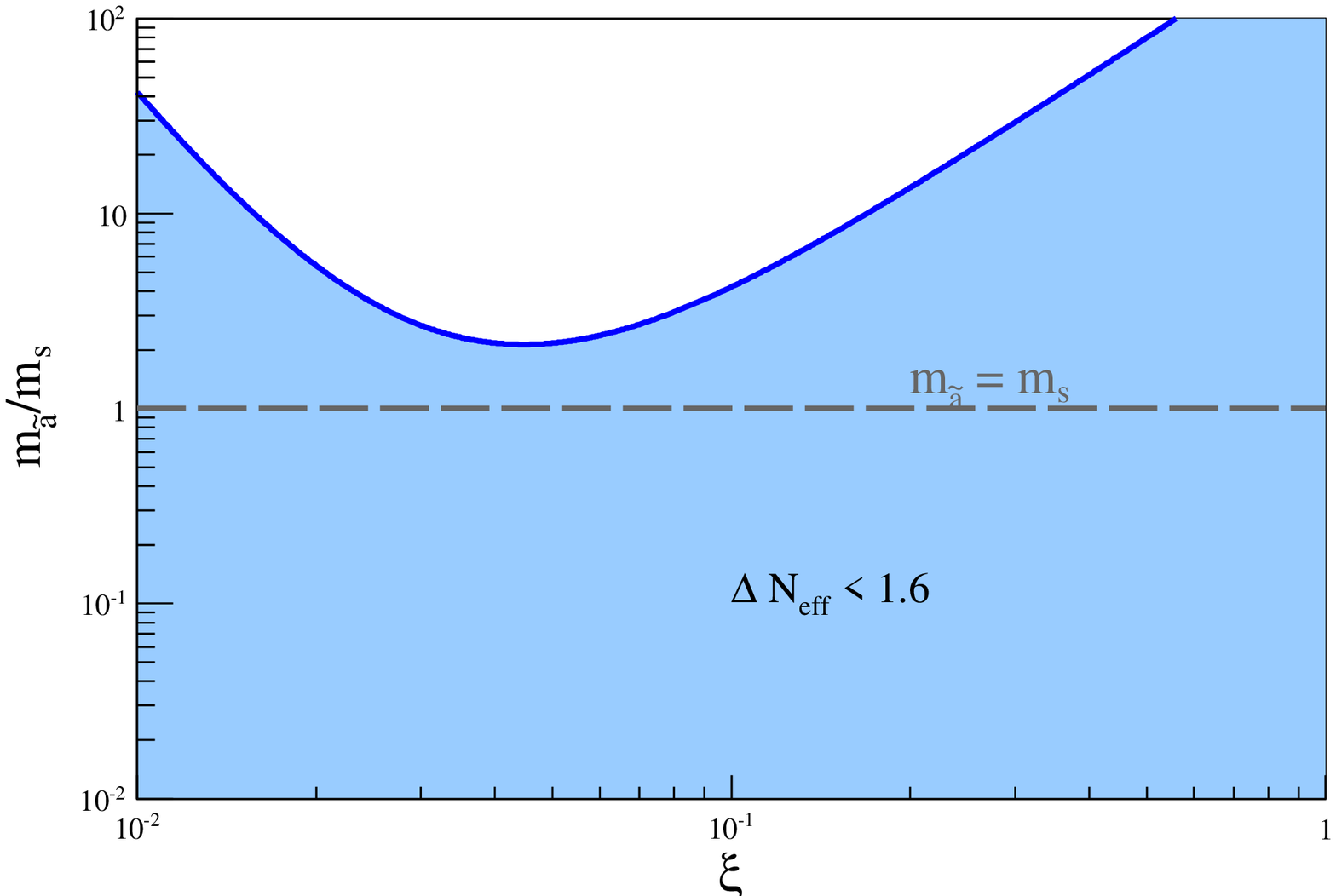}
\caption{The upper bound on $m_{\ta}/m_s$ as a function of $\xi$ according to Eq.~(\ref{eq:ntpmaxr}). 
The shaded region below the curve satisfies $\Delta N_{eff} < 1.6$, {\it irrespective of the other PQ parameter values}.
The region above the curve may be either excluded or allowed by CMB constraints depending on the values of $T_R$, $f_a$, $m_s$ and $m_{\tg}$.}
\label{fig:xibounds}}
%


\subsection{$\Delta N_{eff}$ from Coherent Oscillations}
\label{sec:co}

As shown in the previous section, the contribution to $\Delta N_{eff}$ 
from thermal production of axions and saxions is suppressed unless $m_{\ta} \gg m_{s}$ and/or $\xi \simeq 0.05$. 
In this section, we assume $m_{\ta} \lesssim m_s$ and $\xi \sim 1$, so $\Delta N_{eff}$ 
from the thermal production is negligible 
and the relic density of relativistic axions is dominated by coherent production of saxions and their decay.
Furthermore, we consider the case for $Y_{\ta}^{TP}\ll Y_s^{CO}$ to exclude the dilution from axino decay.
However, even if the axino thermal production is suppressed, axinos can still be non-thermally produced through
saxion decays, if $m_s > 2 m_{\ta}$. In this case both CO saxions and non-thermally produced axinos may inject entropy
in the early universe at different times, what considerably complicates the picture. Such scenarios will be fully
addressed once we present our numerical results in Sec.~\ref{sec:num}. Here, for simplicity, we will assume $m_s < 2 m_{\ta}$,
so both the thermal and non-thermal production of axinos can be safely neglected. At the end of this section we will briefly
discuss what happens if $m_s > 2 m_{\ta}$.

Under the above assumptions Eq.~(\ref{eq:dneff2b}) becomes:
\be
\Delta N_{eff} \simeq 18.02 \frac{1}{r} BR(s\to aa) g_{*S}(T_D)^{-1/3} \frac{m_s Y_s^{CO}}{T_D} \label{eq:dneffco}
\ee
and
\be
r \simeq \max\left[1,\frac{4}{3} R(s \to X) \frac{m_{s} Y_{s}^{CO}}{T_D}\right] .
\label{eq:rco}
\ee


Since the only invisible decay mode of the saxion is $s \to aa$, we can approximate the fraction of visible energy injected from saxion decays by:
\be
R(s \to X)  = 1 - BR(s \to aa) \simeq 1 - \frac{\xi^2}{\xi^2  + \alpha_s^2/\pi^2} = \frac{\alpha_s^2/\pi^2}{\xi^2  + \alpha_s^2/\pi^2}
\ee
where we used the result from Eq.~(\ref{eq:br}) and have once again neglected the decay into axinos, since here we assume $m_s < 2 m_{\ta}$.
Combining Eqs.~(\ref{eq:dneffco}) and (\ref{eq:rco}) we obtain:
\be
\Delta N_{eff} \simeq 18.02 g_{*S}(T_D)^{-1/3} \left\{ \begin{array}{ll} \biggl(\frac{\xi^2}{\xi^2+\alpha_s^2/\pi^2}\biggr) \biggl(\frac{m_s Y_s^{CO}}{T_D}\biggr) & \mbox{ , if $r = 1$}  \\
\frac{3}{4} \biggl(\frac{\xi^2}{\alpha_s^2/\pi^2}\biggr) & \mbox{ , if $r > 1$} \end{array} \right. .
\label{eq:dneffco2}
\ee
Since $g_{*S}(T_D)^{-1/3} \gtrsim 0.1$ and $\alpha_s^2/\pi^2 \simeq 10^{-3}$, 
once $s\to aa$ dominates over the visible mode $s\to gg$, i.e. $\xi\gtrsim0.05$, the above result shows that, for coherent production of saxions, {\it the case of $r > 1$ is automatically
excluded by the CMB constraint}. This means that the universe can never have a saxion dominated era and has always been radiation dominated
until late times ($T \ll 1$ MeV). We stress that this conclusion holds as long as saxions are mainly produced through coherent oscillations ($Y_s^{TP} \ll Y_s^{CO}$) and $m_s < 2m_{\ta}$.

On the other hand, if $r = 1$ (no saxion dominated era and no entropy injection), the constraints on 
$\Delta N_{eff}$ give an upper bound on $Y_s^{CO} m_s/T_D$. 
In order to rewrite this in terms of the PQ parameters, we use Eqs.~(\ref{eq:co}), (\ref{eq:tds}) and (\ref{eq:gammas}), which gives:
\bea
\frac{m_s Y_s^{CO}}{T_D} & \simeq & 2.2\times 10^{-6} \frac{1}{\sqrt{\xi^2 + \alpha_s^2/\pi^2}} g_{*S}(T_D)^{1/4} \left(\frac{\min[T_R,T_s]}{10^{8} \mbox{ GeV}}\right) \nonumber \\
& \times & \left(\frac{10^{3}\ {\rm GeV}}{m_s}\right)^{3/2}\left(\frac{s_0}{10^{12}\ {\rm GeV}}\right)^2 \left(\frac{f_a}{10^{12}\ {\rm GeV}}\right) 
\eea
where $T_s \simeq 1.3 \times 10^{10} \sqrt{m_s/10^3\; {\rm GeV}}$ is the saxion oscillation temperature~\cite{AY}.
Thus, from Eq.~(\ref{eq:dneffco2}) for $r=1$:
\bea
\Delta N_{eff} &\simeq & 2.7 \times 10^{-5} \frac{\xi^2}{(\xi^2+\alpha_s^2/\pi^2)^{3/2}} \left(\frac{\min[T_R,T_s]}{10^8 \mbox{ GeV}}\right) \nonumber \\
&\times &  \left(\frac{10^3 \mbox{ GeV}}{m_s}\right)^{3/2}  \left(\frac{s_0}{10^{12} \mbox{ GeV}}\right)^2\left(\frac{f_a}{10^{12} \mbox{ GeV}}\right) \label{eq:neffco}
\eea
where we took $g_{*S}(T_D) \sim 100$.

To illustrate the above results and check their validity, we show in Fig.~\ref{fig:BM2_omgvxi} the numerical solution for $\Delta N_{eff}$
 computed using the full set of coupled Boltzmann equations. We assume $m_s = 2$ TeV, $m_{\ta} = 1.5$ TeV, $m_{\tilde{g}}=1.6$ TeV, 
$T_R=10^6$ GeV, $s_0=f_a=3\times10^{14}$ GeV, $\theta_i=0.01$ and vary $\xi$. We take the SUA benchmark point.
The dotted gray line shows the entropy dilution factor, $r$, and the dashed lines the analytical solutions from Eqs.~(\ref{eq:neffco}) and (\ref{eq:dneffco2}).
As we can see, for the high $f_a$ chosen, $\Delta N_{eff}$ violates the CMB constraint for $\xi \gtrsim 0.02$.
Also we see that the numerical solution follows the behavior expected from Eq.~(\ref{eq:dneffco2}) for $r>1$ ($\xi \lesssim 0.1$).
For higher values of $\xi$, the $s \to gg$ is extremely suppressed, so there is no entropy injection and the solution follows the behavior described by Eq.~(\ref{eq:neffco}) instead.
The small difference between the analytical and numerical solutions is expected, due to the approximations used in the analytical calculation.
We also show the values of the neutralino and CO axion relic densities. In this case, since the
$s \to \ta\ta$ and $s \to \tg\tg$ decays are kinematically forbidden, there is no neutralino injection from saxion decays and its relic density depends on
$\xi$ only through the entropy dilution factor. As we can see both $\Omega_a^{CO}h^2$ and $\Omega_{\tilde{Z}_1}h^2$ scale as $1/r$, as expected.

\FIGURE[t]{
\includegraphics[width=12cm,clip]{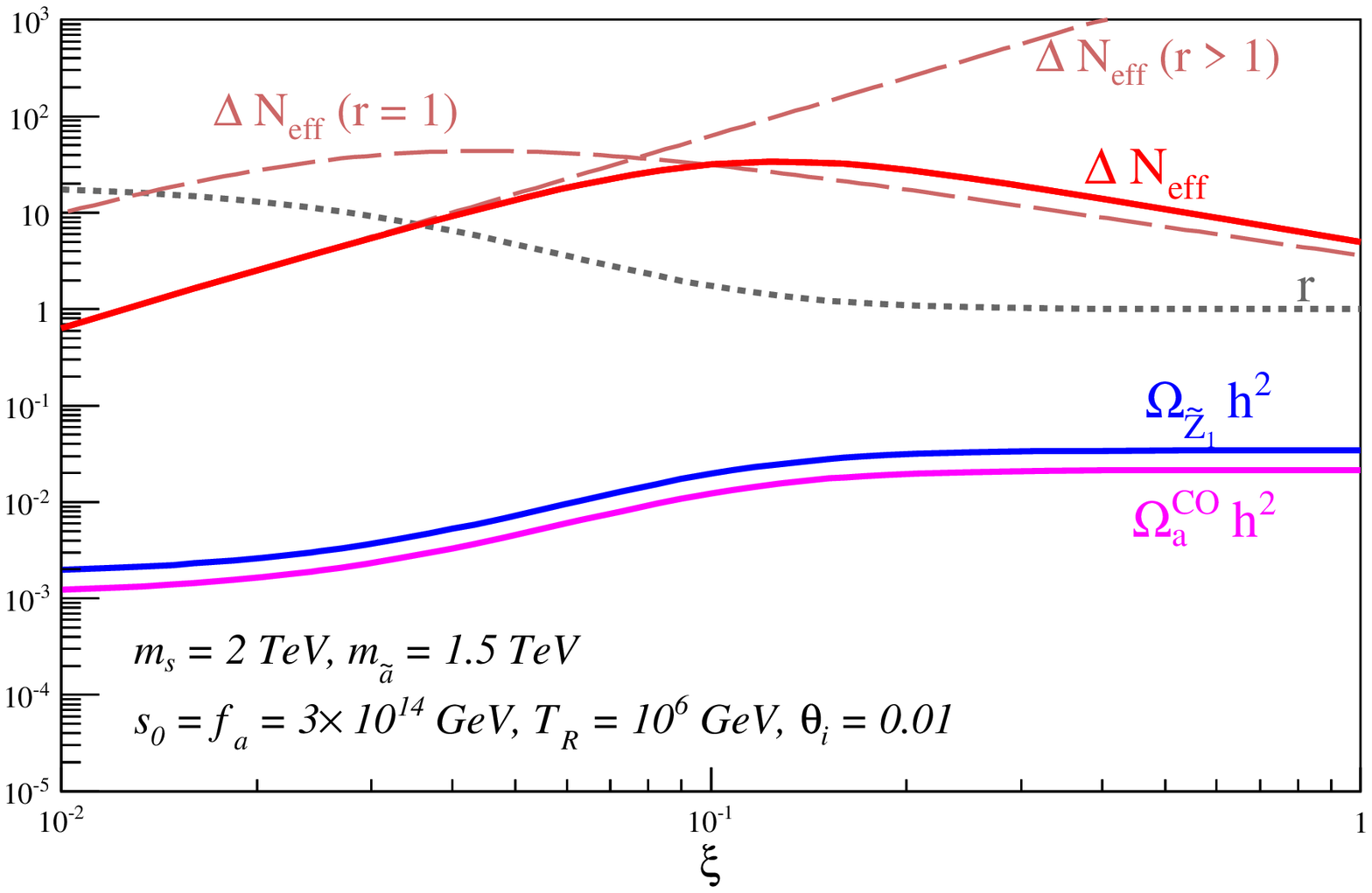}
\caption{Numerical solutions for $\Delta N_{eff}$, $\Omega_a^{CO}h^2$ and $\Omega_{\tilde{Z}_1}h^2$ as a function of $\xi$ for 
CO-produced saxion case. The dotted gray lines show the entropy dilution factor, $r$, while the dashed lines show the analytical results from
by Eqs.~(\ref{eq:neffco}) and (\ref{eq:dneffco2}).
We consider the parameters $m_s = 2$ TeV, $m_{\ta} = 1.5$ TeV, $m_{\tilde{g}}=1.6$ TeV, $T_R=10^6$ GeV, $s_0=f_a=3\times10^{14}$ GeV and $\theta_i=0.01$.
The MSSM point assumed is the SUA listed in Table~\ref{tab:bm}.}
\label{fig:BM2_omgvxi}}

The impact of the CMB constraints on the PQ parameter space for $\xi=1$ is summarized in Fig.~\ref{fig:fabounds}, 
where we show the excluded region in the $f_a$-$m_s$ plane for different values of $T_R$ and $s_0$.
The constrained region is computed using the analytical results of Eq.~(\ref{eq:neffco}).
For $s_0 = f_a$ and $m_s \sim 1$ TeV, the CMB
constraint on dark radiation requires $f_a \lesssim 10^{12}-10^{14}$ GeV depending on $T_R$.
On the other hand, if the saxion field amplitude $s_0$ is $\sim M_P/100$ (as suggested in some models~\cite{kkn} 
and assumed in Fig.~\ref{fig:fabounds}{\it b})
we see that relativistic axions from saxion decays easily violate the CMB constraint, unless $m_s \gtrsim 200$ GeV, $f_a \lesssim 10^{13}$ GeV and $T_R \lesssim 10^6$ GeV.
As we can see, the constraint $\Delta N_{eff} < 1.6$ imposes an {\it upper bound} on $f_a$, 
which strongly depends on the value of $s_0$, since this parameter controls the 
amplitude of saxion coherent oscillations. 
We point out that this constraint is independent of the misalignment angle $\theta_i$
and is not related to the traditional upper bound on $f_a$ ($\lesssim \theta_i^{-2}\, 10^{12}$ GeV) coming from the overclosure of the universe from CO production
of axions. Thus the dark radiation constraint provides an {\it additional} and independent constraint on $f_a$.
We also point out that, although in  Fig.~\ref{fig:fabounds} we have assumed $\xi=1$, we expect even stronger constraints for $\xi \lesssim 1$, since $\Delta N_{eff}$ increases
as $\xi$ decreases (as long as $\xi \gtrsim 0.05$), as shown in Fig.~\ref{fig:BM2_omgvxi}.

\FIGURE[t]{
\includegraphics[width=7.3cm,clip]{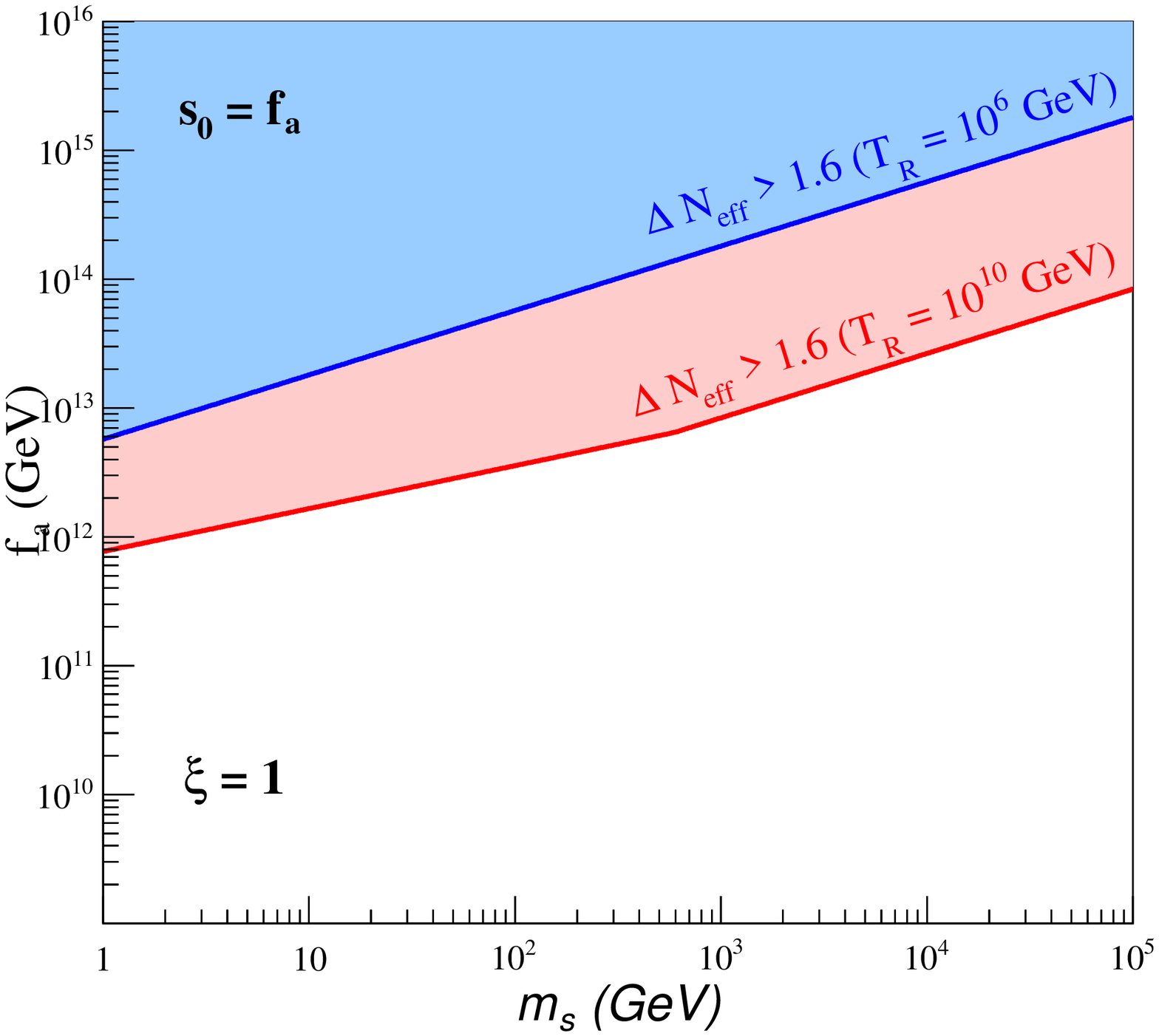}
\includegraphics[width=7.3cm,clip]{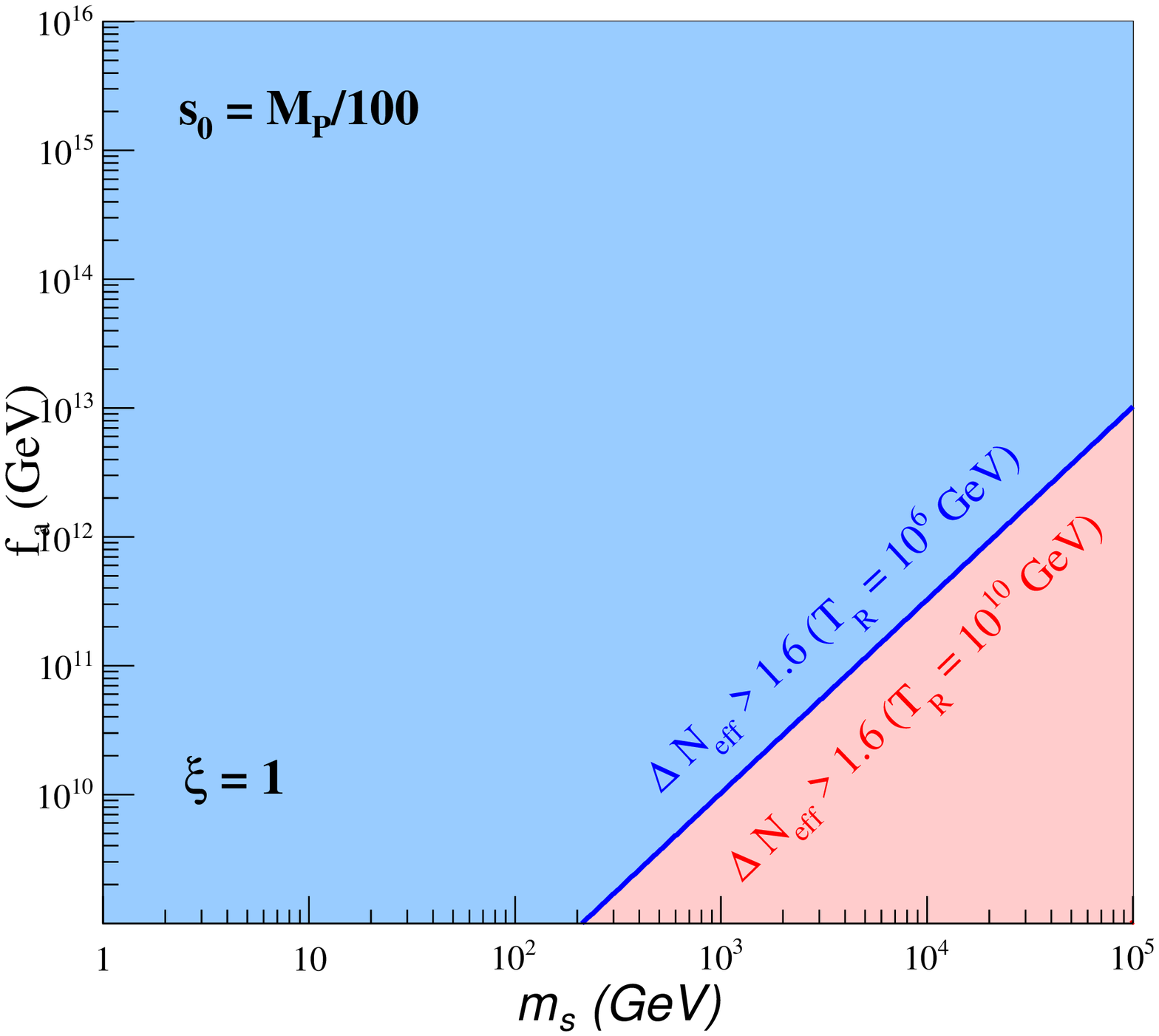}
\caption{Dark radiation bounds on the $f_a$-$m_s$ plane for $s_0 = f_a$ (left) and $s_0 = M_P/100$ (right).
The region above the solid red (blue) line has $\Delta N_{eff} > 1.6$ for $T_R = 10^{10}$ GeV ($T_R = 10^{6}$ GeV).
The curves assume $m_s < 2 m_{\ta}$, only include the contribution from CO saxions and were obtained using Eq.~(\ref{eq:neffco})
and $\xi =1$.}
\label{fig:fabounds}}

So far all the results presented in this section have assumed $m_s < 2m_{\ta}$.
As mentioned before, if the $s \to \ta \ta$ decay is kinematically allowed, saxion decays to axinos
can result in an even later entropy and  neutralino injections (through the axino cascade decay), even if
thermal production of axinos is suppressed. 
Since in this case there are two phases of entropy injection (at the saxion and axino decays)
and the universe can go through saxion and axino dominated eras, it becomes difficult to treat it analytically.
Therefore, to discuss the $m_s > 2m_{\ta}$ case we use the numerical method discussed in the Appendix.
We show in Fig.~\ref{fig:BM2_omgvms}a the numerical solutions for 
$\Delta N_{eff}$, $\Omega_a^{CO}h^2$, $\Omega_{\tilde{Z}_1}h^2$ and $r$ as functions of $m_s$.
We take $m_{\tilde{g}}=1.6$ TeV, $T_R=10^6$ GeV, $s_0=f_a=10^{15}$ GeV, $\theta_i=0.01$ and $\xi=1$.
For $m_s < m_{\ta}$ we fall into the scenario discussed above and since we have $r>1$ in this region, 
$\Delta N_{eff}$ is approximately constant, as expected from Eq.~(\ref{eq:dneffco2}). Also $\Delta N_{eff} \gg 1$,
as anticipated by our previous results for the case $r>1$. However, once $m_s > 2m_{\ta}$, the $s \to \ta\ta$
decay becomes kinematically allowed and drastically reduces  $\Delta N_{eff}$. This is mainly due to two reasons.
First, as seen in Fig.~\ref{fig:sBF}, $BR(s \to aa)$ decreases around $m_s \gtrsim 2m_{\ta}$, thus
decreasing the injection of axions from saxion decays. Second, the injection of axinos and their subsequent cascade decay
to neutralinos significantly injects entropy at later times, thus suppressing $\Delta N_{eff}$. The sudden
enhancement in entropy injection once $s \to \ta\ta$ opens up is shown by the rapid increase in $r$ around $m_s = 1$ TeV.

To illustrate these effects, we show in Fig.~\ref{fig:BM2_omgvms}b
the cosmological evolution of the energy densities of radiation, neutralinos, axions, saxions and axinos as a function
of the scale factor $R$. The PQ parameters are the same used in Fig.~\ref{fig:BM2_omgvms}a and $m_s = 2$ TeV.
As we can see, as saxions start to decay (around $R/R_0 \sim 10^3$), the energy density of axions and axinos rapidly increases until $R/R_0 \sim 10^9$, where
the universe goes from a saxion dominated to an axino dominated era. At much later times ($R/R_0 \sim 10^{13}$) the axino decays
and injects entropy, significantly diluting the relic density of both relativistic and cold CO axions.
As a result, $\Delta N_{eff}$ and  $\Omega_a^{CO}h^2$  become highly suppressed, as seen on Fig.~\ref{fig:BM2_omgvms}a.
Therefore, due to the late entropy injection from axino decays, the CMB
constraint on $\Delta N_{eff}$ can be easily satisfied even for a $f_a$ value well above the bounds from Fig.~\ref{fig:fabounds}.
However, the injection of neutralinos from non-thermally produced axinos easily
surpass the observed DM relic abundance, as shown in Fig.~\ref{fig:BM2_omgvms}.
As a consequence, it seems difficult to simultaneously satisfy the CMB constraints on  $\Delta N_{eff}$ and $\Omega_{DM} h^2$
for large $f_a$. In the next section we will generalize this result for more arbitrary choices of the PQ parameters.

\FIGURE[t]{
\includegraphics[width=7.cm,clip]{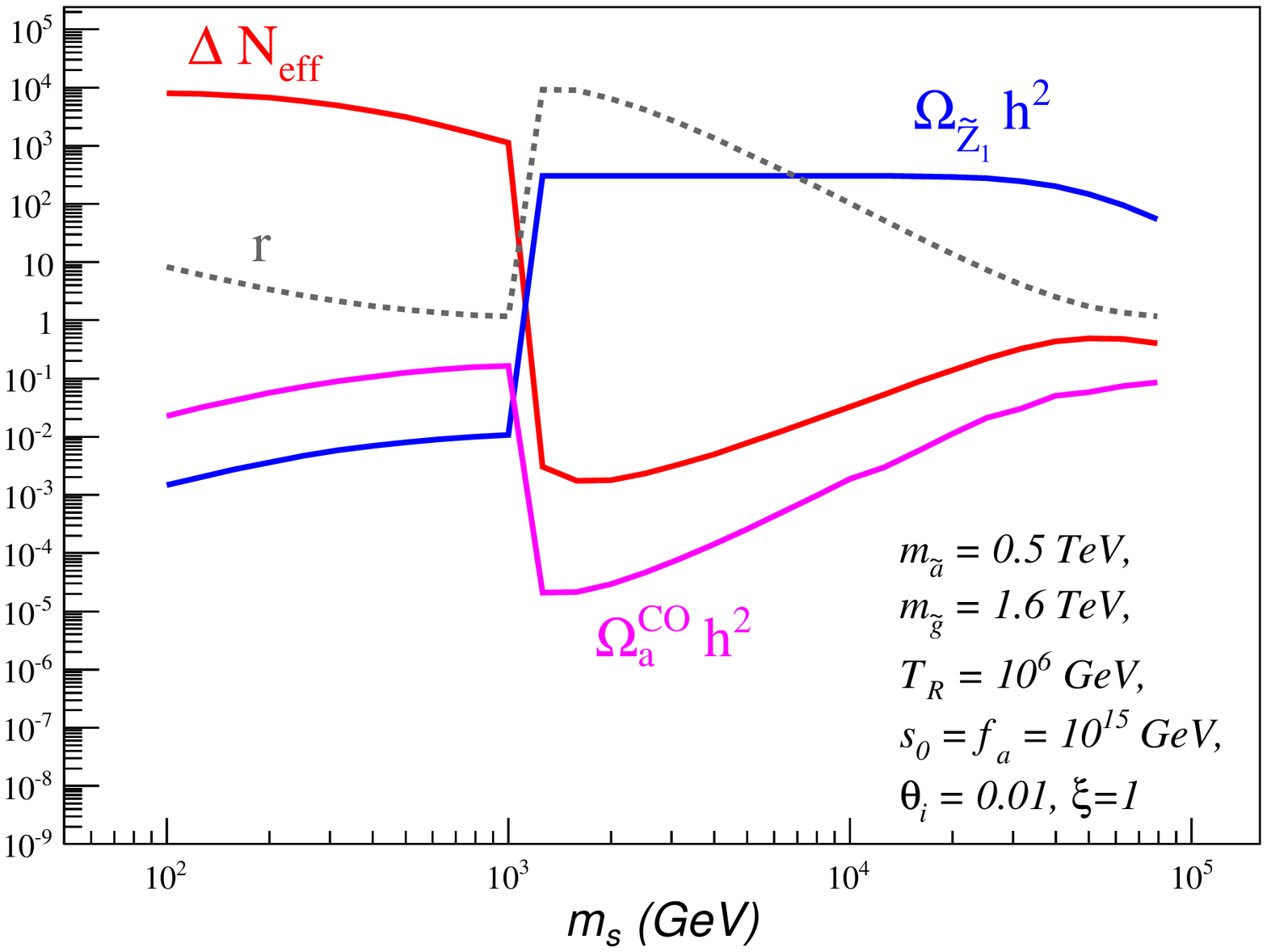}
\includegraphics[width=7.cm,clip]{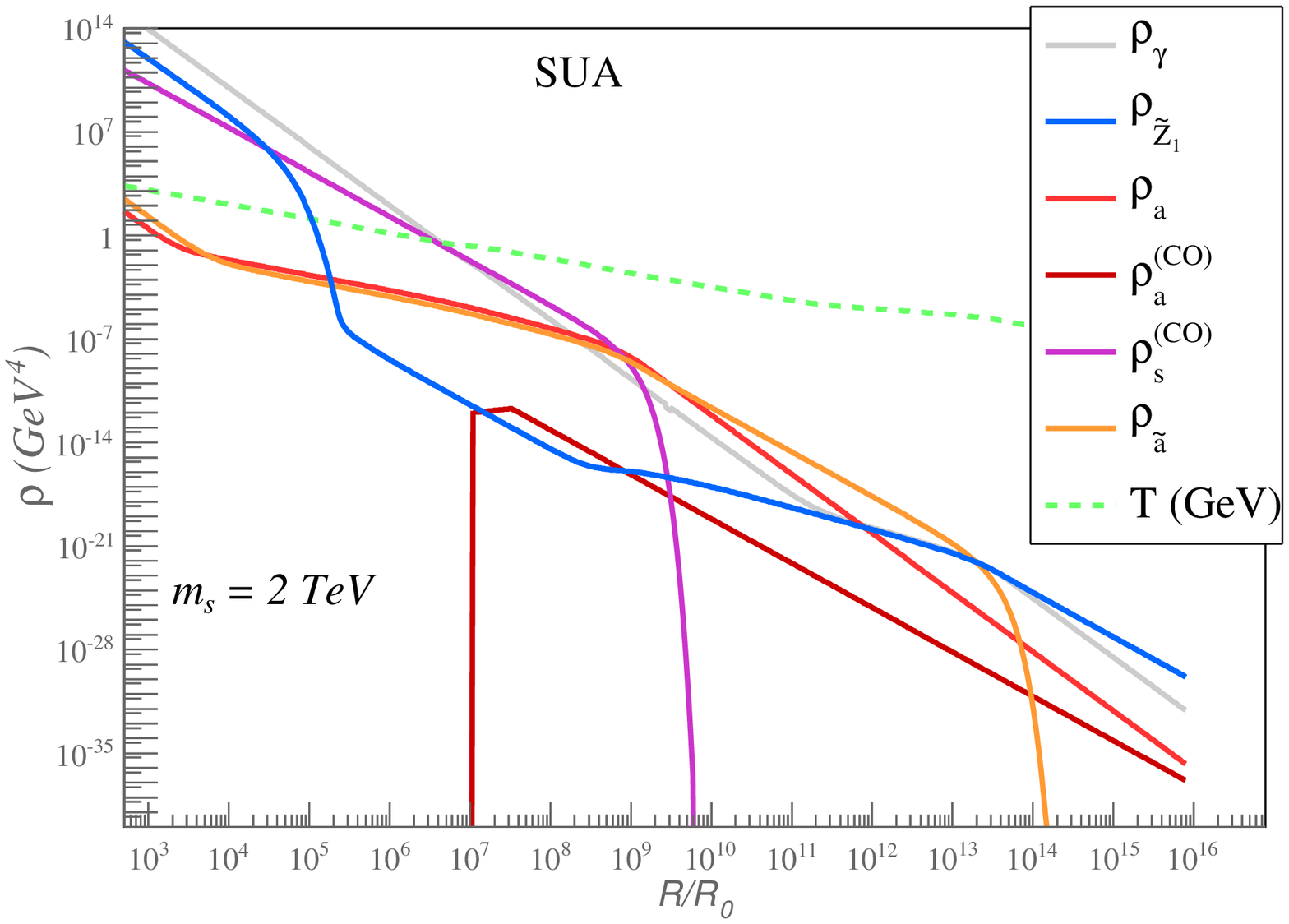}
\caption{Left: $\Delta N_{eff}$, $\Omega_a^{CO}h^2$, $\Omega_{\tilde{Z}_1}h^2$ and $r$ as functions of $m_s$ for the
CO-produced saxion case. We consider the PQ parameters $T_R=10^6$ GeV, $s_0=f_a=10^{15}$ GeV and $\theta_i=0.01$ and the SUA point. 
Right: energy densities versus the scale factor for the same PQ parameters and $m_s=2$ TeV.}
\label{fig:BM2_omgvms}}

\subsection{Numerical Results}
\label{sec:num}

In the previous sections, we have discussed some particular scenarios and 
derived useful analytical approximations for $\Delta N_{eff}$. 
Here we generalize these results using the full numerical solutions for the coupled Boltzmann equations, 
simultaneously including all production mechanisms (thermal scatterings and coherent oscillations) for the PQ fields.
We will also discuss the DM content of the viable scenarios.
To keep our results as general as possible, we scan over the following PQ parameter values:
\bea
10^9\ {\rm GeV} < & f_a & < 10^{16}\ {\rm GeV},\nonumber \\
0.3\ {\rm TeV}  < & m_{\ta} & < 20 \ {\rm TeV},\nonumber \\
0.3\ {\rm TeV} < & m_s & < 20\ {\rm TeV},\nonumber \\
10^{-4} < & s_0/f_a & < 10^{4},\nonumber \\
10^6\ {\rm GeV} < & T_R & < \min(f_a,10^{10}\ {\rm GeV}) \label{eq:scan}
\eea
while keeping $\xi = 1$ and $m_{\tG} = 3$ TeV fixed as well as the MSSM spectrum. 
For each point in parameter space, we use the numerical solutions
of the Boltzmann equations to compute $\Delta N_{eff}$ and other quantities of interest.

In Fig.~\ref{fig:scan1} we once again show $\Delta N_{eff}$ {\it vs.} $f_a$, 
but now varying all the PQ parameters within the parameter space defined in Eq.~(\ref{eq:scan}) 
and using the SUA parameters for MSSM spectrum. 
In order to compare the full results to the analytical approximations of the previous sections, 
we show in different colors points with $s_{0}/f_{a} =10^{-4}$ (blue), $s_{0}/f_{a} \leq 1$ (magenta) and
$s_{0}/f_{a} > 1$ (red). 
For the blue points, the low $s_{0}/f_{a}$ value suppresses the coherent production of saxions,
so the main contribution to $\Delta N_{eff}$ comes from thermal production, except for very high 
$f_a$ values ($\gtrsim 10^{14}$ GeV).
These points are described by the results from Sec.~\ref{sec:tp} and the blue dashed line shows the 
maximum $\Delta N_{eff}$ allowed by Eq.~(\ref{eq:ntpmaxr}) ($\simeq 0.8$, for $m_{\ta}/m_s < 20/0.3$ and $\xi = 1$).
As we can see, this upper bound is respected by the numerical solutions 
even when all the PQ parameters are varied, except for the cases where there is a signficant contribution
from coherent production of saxions (magenta and red points). 

From Fig.~\ref{fig:scan1}, we see that the conclusion from Sec.~\ref{sec:tp} is preserved
even in a more general scan: {\it thermal production of saxions can easily satisfy the CMB constraint}.
However, once production of saxions through coherent oscillations is included (represented by
magenta and red points at low $f_a$ and by all points at large $f_a$), 
large values of $\Delta N_{eff}$ can be generated. Nonetheless we find that for sufficiently low $T_R$
and heavy saxions, values of $f_a$ as large as $10^{16}$ GeV can still be consistent with the CMB constraint
even for $s_0 \geq f_a$. All these solutions have $m_s > 2m_{\ta}$ and correpond to the cases where entropy injection from axino
decays highly suppress $\Delta N_{eff}$, as shown by the examples in Fig.~\ref{fig:BM2_omgvms}.
%
\FIGURE[t]{
\includegraphics[width=13cm,clip]{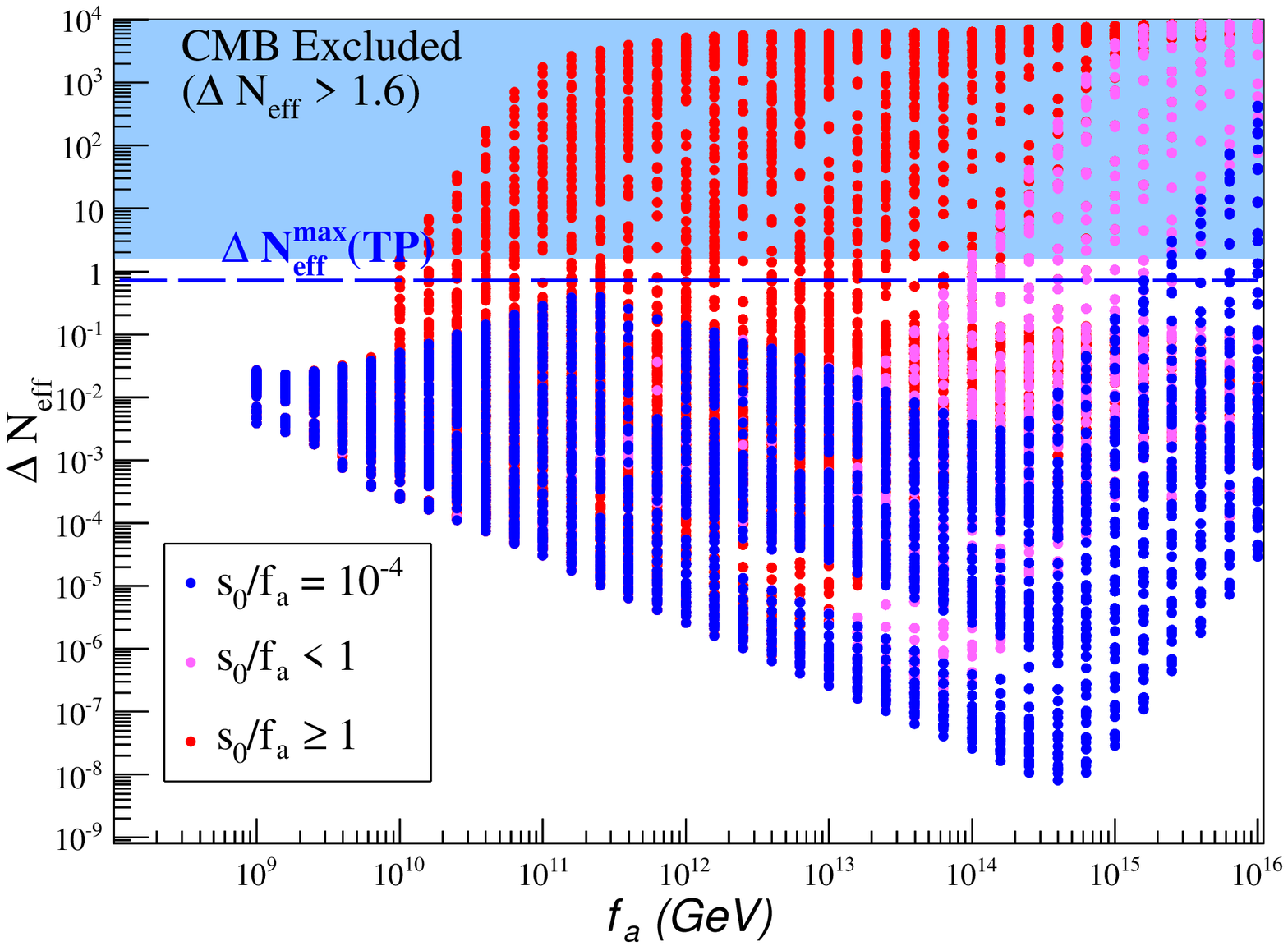}
\caption{$\Delta N_{eff}$ as a function of the PQ breaking scale $f_a$ for the scan over the PQ parameter space
defined in Eq.~(\ref{eq:scan}) for the SUA benchmark. 
Blue points have $s_0/f_a = 10^{-4}$ and 
are dominated by thermal saxion production for $f_a \lesssim 10^{14}$ GeV. 
Purple points have $s_0/f_a < 1$ and red points have $s_0/f_a \geq 1$.
The dashed blue line shows the maximum expected value for $\Delta N_{eff}$ from thermal saxion production, 
as obtained from Eq.~(\ref{eq:ntpmaxr}).
The shaded region violates the CMB constraint on dark radiation.}
\label{fig:scan1}}

In Fig.~\ref{fig:scan2}, we show the neutralino relic density as a function of $f_a$
again for the benchmark point SUA,
which has an standard underabundance of neutralino DM due to a higgsino-like neutralino. 
Blue (red) points are allowed (excluded) by BBN constraints and have $\Delta N_{eff} < 1.6$, 
while magenta points have $\Delta N_{eff} > 1.6$. The green points are both allowed by BBN and lie in the
$1\sigma$ interval for $\Delta N_{eff}$ from the current WMAP9 results.
The standard thermal value for $\Omega_{\tz_1} h^2$ is shown by the dashed gray line and 
we see that for $f_a \lesssim 10^{13}$ GeV the neutralino relic abundance is enhanced by 
TP axino decays, $s\to \ta \ta$ and/or $s\to\tg\tg$ decays. 
For larger values of $f_a$, there are several solutions with suppressed values of 
$\Omega_{\tz_1} h^2$ when compared to the MSSM value.
These points usually have suppressed axino and thermal saxion production 
(due to the large $f_a$ value) and $s \to \tg \tg$ is forbidden ($m_s<2m_{\tg}$). 
In this case, the injection of neutralinos from axino and saxion decays
is highly suppressed and easily compensated by the entropy injection from CO-produced 
saxions followed by decays to gluons. 
However, as shown in Fig.~\ref{fig:scan2},
all these points have too large values of $\Delta N_{eff}$. 
This is in agreement with the results of Sec.~\ref{sec:co},
where we showed that it is not possible to have entropy dilution ($r > 1$) 
from coherent oscillating saxions without either violating the CMB constraint on dark radiation or overclosing the universe ($\Omega_{\tz_1} h^2 \gg 0.11$).
The blue-shaded region in the Figure is excluded by applying the recent Xe-100 WIMP search
bounds~\cite{xe100} to SUA with a re-scaled local abundance of WIMPs. As we can see, for the SUA point, the large annihilation cross-section required to suppress
the neutralino relic is related to a high $\sigma^{SI}(\tz_1 p)$, hence this point is subject to stringent bounds from Xe-100.

Therefore, we conclude that in order to have $\Delta N_{eff} < 1.6$,
{\it the neutralino abundance can only be enhanced with respect to its thermal value}.
Hence only SUSY models with a standard {\it underabundance} of relic neutralinos
can be consistent with the simultaneous constraints on the dark matter density, $\Delta N_{eff}$ and 
the constraints from BBN. 
This result is confirmed by Fig.~\ref{fig:scan3} where we again show 
the neutralino relic density as a function of $f_a$, but now for the SOA benchmark, which has a bino-like neutralino
with a standard overabundance ($\Omega_{\tz_1}^{MSSM} h^2 = 6.8$). As we can see, all points consistent
with the observed CDM abundance are excluded by the $\Delta N_{eff} < 1.6$ constraint.

\FIGURE[t]{
\includegraphics[width=13cm,clip]{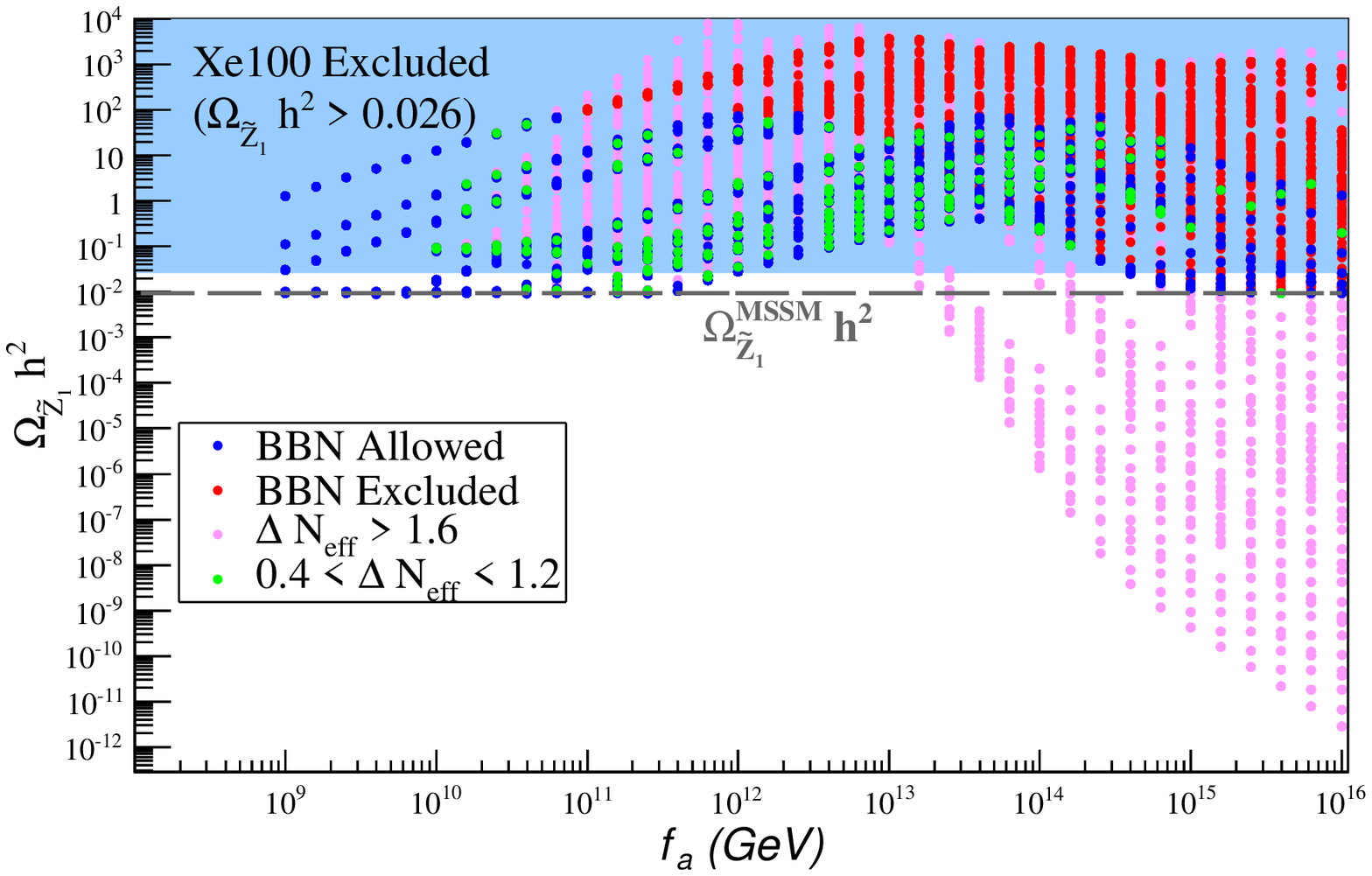}
\caption{$\Omega_{\tz_1} h^2$ as a function of the PQ breaking scale $f_a$
for the scan over the PQ parameter space defined in Eq.~(\ref{eq:scan}), assuming
the benchmark point SUA, listed in Table~\ref{tab:bm}.
Blue and red points have $\Delta N_{eff} < 1.6$, while green points have $0.4 < \Delta N_{eff} < 1.2$  and magenta points have 
$\Delta N_{eff} > 0.1.6$.
Also, blue and green points are allowed by the BBN constraints on decaying saxions, axinos and 
gravitinos, while red points are excluded. 
The gray dashed line shows the standard thermal value $\Omega_{\tz_1}^{TP} h^2$ in the MSSM.
The blue-shaded region is excluded by Xe-100 WIMP searches at $m_{\tz_1}=135.4$ GeV
after applying a re-scaled local WIMP abundance.
}
\label{fig:scan2}}
\FIGURE[t]{
\includegraphics[width=13cm,clip]{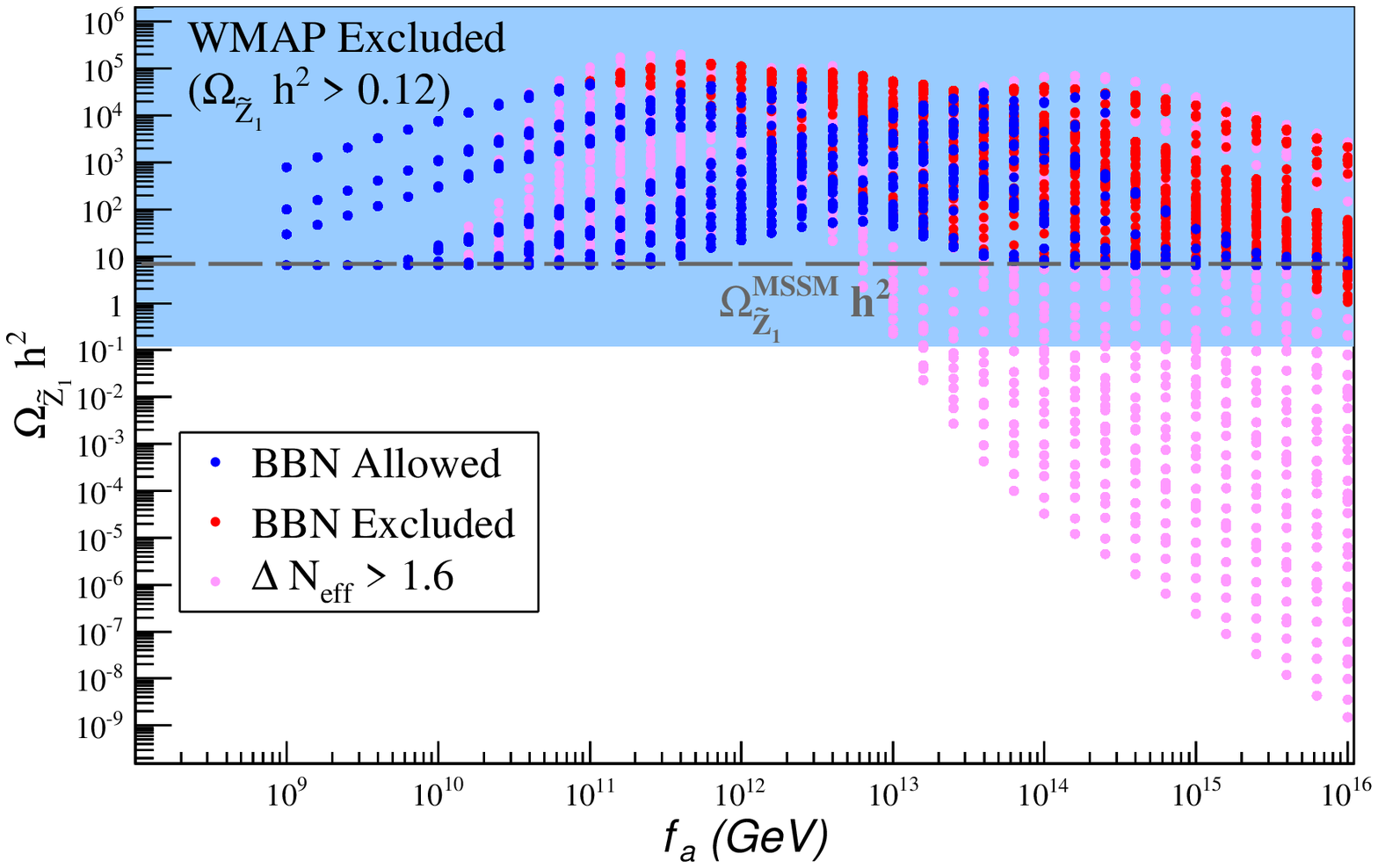}
\caption{Same as in Fig.~\ref{fig:scan2} but for the benchmark point SOA.}
\label{fig:scan3}}

Finally, we comment on the possibility of the excess in $N_{eff}$ seen by the WMAP9 and
SPT analyses being real. In this case, as shown by the results from Secs.~\ref{sec:tp} and ~\ref{sec:co}, such an excess
can only be generated by CO-production of saxions, unless $m_{\ta} \gg m_s$. Also, as shown by the results in Fig.~\ref{fig:BM2_omgvms},
$\Delta N_{eff} \sim 1$ can only be obtained if the saxion decay to axinos is suppressed ($m_s < 2m_{\ta}$), otherwise entropy injection from NTP axinos
efficiently dilute $\Delta N_{eff}$ to extremely low values.
Furthermore, as shown by Fig.~\ref{fig:scan1}, $\Delta N_{eff} \sim 1$ can be obtained for a wide range of $f_a$ values, depending on the value of $s_0$.
This is also seen in Fig.~\ref{fig:scan2}, where we show as green
points the solutions which simultaneously satisfy the BBN constraints and lie in the $1\sigma$ interval for $\Delta N_{eff}$
given by the WMAP9 analysis. 
As we can see, $\Delta N_{eff} \sim 1$ can be easily obtained as long as $f_a \gtrsim 10^{10}$ GeV,
where CO-production of saxions becomes sufficiently large to generate the excess. We point out that this lower bound on $f_a$
is directly related to the maximum value of $s_0$ ($\le 10^4 f_a$) and the $\xi$ value assumed in our scan. For higher values of $s_0$ and
smaller values of $\xi$ ($<1$), lower $f_a$ values would be consistent with $\Delta N_{eff} \sim 1$.
From Fig.~\ref{fig:scan2} we also see that the allowed range of $f_a$ is more strongly constrained by DM and Xe100
bounds than by the $\Delta N_{eff} \sim 1$ requirement.
Therefore we conclude that, {\it if $\Delta N_{eff} \simeq 1$ is eventually confirmed by Planck data} and $m_{\tz_1} < m_{\ta} \lesssim m_s$ is assumed,
saxions were coherently produced in the early universe at large rates and $m_s < 2m_{\ta}$.

\section{Conclusions}
\label{sec:conclude}

In this paper, we have presented the impact of the CMB constraints on dark radiation
for the mixed neutralino/axion dark matter scenario. To properly compute the dark
matter and dark radiation relic abundances we have made use of a simplified set of
coupled Boltzmann equations, as described in the Appendix.

We discussed the case of large $\xi$ ($\gtrsim 0.05$) where saxion decays
to $aa$ and $\ta\ta$ are dominant. The case of small $\xi$, where saxions
dominantly decay to $gg$ or $\tg\tg$ was
previously presented in Ref.~\cite{bls}.
In the present case, $s\to aa$ may contribute to dark radiation with
a contribution parametrized as $\Delta N_{eff}$, the non-standard
contribution to the number of effective neutrinos. 
Recent CMB 
analyses restrict $\Delta N_{eff}<1.6$, providing a strong constraint
on models producing dark radiation. 

Our main results may be summarized as follows.
\begin{itemize}
\item At low $f_a$ and $s_0$ ($\sim 10^9-10^{11}$ GeV), 
saxion production is expected to be dominated
by thermal production. In this case, the $\Delta N_{eff}<1.6$ bound
only weakly constrains the PQMSSM parameter space as summarized in Fig.~\ref{fig:xibounds}. 
\item At high  $f_a$ and $s_0$ ($\sim 10^{12}-10^{16}$ GeV), thermal production of saxions, axinos and axions are suppressed
and coherent production of saxions dominates. In this case, the constraints on dark radiation
provide an upper bound for $f_a$ and $s_0$, as shown in Fig.~\ref{fig:fabounds}.
\end{itemize}

Once the CMB constraints on $\Delta N_{eff}$ are combined with the BBN constraints on late decaying particles
and the constraints on the dark matter relic abundance, we have found that:
\begin{itemize}
\item In the case of SUSY models with a standard overabundance of neutralinos
(as in our SOA benchmark case), it is possible to dilute the relic abundance of neutralinos below the observed
DM relic abundance through entropy injection from saxion decays to gluons. However such solutions always violate
the CMB bound on $\Delta N_{eff}$.
Therefore we find that SUSY models with a SOA of neutralinos {\it are still excluded for all choices of PQ parameters}.
\item In the case of SUSY models with a SUA of neutralinos, 
low values of $f_a$ ($\sim 10^9-10^{12}$) can easily accommodate all the constraints, since in this
case saxions and axinos are mainly thermally produced and are short-lived, thus suppressing
their contributions to $\Delta N_{eff}$ and  $\Omega_{\tz_1}h^2$.
Once $f_a \gtrsim 10^{12}$ GeV, axinos become long-lived and enhance the neutralino abundance.
In most cases, the augmentation leads to an overabundance of
neutralinos. However, for very high values of $f_a$ ($\gtrsim 10^{15}$ GeV) and small values of $s_0$ ($\lesssim 10^{-3}$),
both the thermal production of saxions and axinos and the production of saxions via coherent oscillations
become suppressed, thus resulting effectively in the usual thermal production of neutralinos 
accompanied by CO-produced axions. Then, once again, the DM, BBN and $\Delta N_{eff}$
constraints can be simultaneously satisfied.
If $\Omega_{\tz_1}h^2$ is less than the observed dark matter relic abundance, 
the remaining DM abundance may be comprised of CO-produced axions, once the appropriate value of the misalignment angle $\theta_i$ is chosen.
These SUA SUSY models tend to
have large enough direct/indirect WIMP detection rates that they should be
soon seen by such experiments.
\end{itemize}

Finally, we point out that if an excess on $N_{eff}$ is confirmed by Planck data, it can be explained
by a significant production of saxions via coherent oscillations followed by $s \rightarrow aa$ decays.
However, saxion decays to axinos must be suppressed in order to avoid subsequent entropy injection from
 axino decays. This is naturally satisfied if $m_s < 2 m_{\ta}$.

{\it Note added}: After completion of this work, we noticed
Ref. \cite{graf} appeared on a similar topic.

\acknowledgments

This research was supported in part by the U.S. Department of Energy
and FAPESP.

\appendix
\section{Boltzmann equations for number and energy densities}
\label{sec:appendix}

The general Boltzmann equation for the number distribution of a particle species can be written as~\cite{kt} 
(assuming isotropy):
\be
\drv{F_{i}}{t} -H p \drv{F_{i}}{p} = C_{i}[F_{i},F_{j},p] \label{eq:d1}
\ee
where $F_{i}(p)$ is the number distribution of particle $i$ as function of momentum $p$, $C$ represents a source/sink term and $H$ is the Hubble constant:
\be
H = \sqrt{\frac{1}{3} \frac{\rho_T}{M_P^2}} 
\label{H} 
\ee 
with $\rho_T = \sum_{i} \rho_i$. The number, energy and pressure densities are given in terms
of $F_{i}$ as:
\bea
n_{i}(t) & = & \int \frac{dp}{2 \pi^2} p^2 F_i(p,t) \nonumber \\ 
\rho_{i}(t) & = & \int \frac{dp}{2 \pi^2} p^2 E_i F_i(p,t) \label{beqs}\\
P_{i}(t) & = & \frac{1}{3} \int \frac{dp}{2 \pi^2} \frac{p^4}{E_i} F_i(p,t) \nonumber
\eea
where $m_i$ is the mass of particle $i$ and $E_i = \sqrt{p_i^2 + m_i^2}$. Using Eq.~(\ref{eq:d1}) 
we obtain the following equations for the number and energy densities:
\bea
\Drv{n_i}{t} + 3H n_i & = & \int \frac{dp}{2 \pi^2} p^2 C_i \nonumber \\
\Drv{\rho_i}{t} + 3H (\rho_i + P_i) & = & \int \frac{dp}{2 \pi^2} p^2 E_i C_i .
\label{eq:meqs}
\eea

We will assume that $C$ is given by:
\be
C = C_{dec} + C_{coll}
\ee
where $C_{dec}$ contains the contributions from decays ($i \to j + X$ and $j \to i + X$) and $C_{coll}$ from
collisions with the thermal plasma. Below we compute each term separately, under some simplifying assumptions.

\subsection{Collision Term}

The collision term $C_{coll}$ for the $i + j \leftrightarrow a + b$ process is given by~\cite{kt}:
\bea
\int \frac{dp}{2 \pi^2} p^2 C_{coll} & = -\int d\Pi_{i} d\Pi_{j} d\Pi_{a} d\Pi_{b} (2 \pi)^4 \delta^{(4)}(p_i + p_j - p_a - p_b) |M|^2 \nonumber \\
& \times \left[ F_i F_j (1 \pm F_a)(1 \pm F_b) - F_a F_b (1 \pm F_i)(1 \pm F_j) \right] 
\eea
where $d\Pi_{i} = d^{3} p_i/((2\pi)^3 2 E_i)$. 
Since we are ultimately interested in Eqs.~(\ref{eq:meqs}) for the number and energy densities, we will
consider the following integral:
\bea
\int \frac{dp}{2 \pi^2} p^2 C_{coll} \times E_i^{\alpha} & = -\int d\Pi_{i} d\Pi_{j} d\Pi_{a} d\Pi_{b} (2 \pi)^4 \delta^{(4)}(p_i + p_j - p_a - p_b) |M|^2 \nonumber \\
& \times \left[ F_i F_j (1 \pm F_a)(1 \pm F_b) - F_a F_b (1 \pm F_i)(1 \pm F_j) \right] \times E_i^{\alpha}
\eea
where $\alpha = 0 (1)$ for the contribution to the number (energy) density equation and the plus (minus) sign is for bosons (fermions). Below we derive a simplified
version of the above equation, valid under some approximations.
The first approximation assumes $F_i \ll 1$, so $1 \pm F_i \simeq 1$. Furthermore, here we assume that the distributions can be
approximated by\footnote{This approximation is only valid for particles with a thermal distribution. However, since the collision term
is responsible for keeping the particle $i$ in thermal equilibrium with the plasma, it is reasonable to assume a thermal distribution for $i$
while the collision term is relevant.}:
\be
F_i \simeq \exp(-(E_i - \mu_i)/T)
\ee
so the collision term can then be written as:
\bea
& \int & \frac{dp}{2 \pi^2} p^2 C_{coll}  E_i^{\alpha} =  -\left( \exp((\mu_i + \mu_j)/T) -\exp((\mu_a + \mu_b)/T)\right) \nonumber \\
 & \times & \int  d\Pi_{i} d\Pi_{j} d\Pi_{a} d\Pi_{b} (2 \pi)^4 \delta^{(4)}(p_i + p_j - p_a - p_b) |M|^2 \exp(-(E_i + E_j)/T) \times E_i^{\alpha} \nonumber
\eea
where above we have used conservation of energy ($E_i + E_j = E_a + E_b$). Since for the cases of interest
the equilibrium distributions have zero chemical potential, we have:
\be
\frac{n_i}{\bar{n}_i} = \exp(\mu_i/T)
\ee
so:
\bea
& \int & \frac{dp}{2 \pi^2} p^2 C_{coll} E_i^{\alpha} = -\left( \frac{n_i n_j}{\bar{n}_i \bar{n}_j} - \frac{n_a n_b}{\bar{n}_a \bar{n}_b}\right) \nonumber \\
 & \times & \int  d\Pi_{i} d\Pi_{j} d\Pi_{a} d\Pi_{b} (2 \pi)^4 \delta^{(4)}(p_i + p_j - p_a - p_b) |M|^2 \exp(-(E_i + E_j)/T) \times E_i^{\alpha} . \nonumber
\eea
In particular, for the process $i + i \leftrightarrow a + b$, where $a$ and $b$ are in thermal equilibrium ($\mu_a = \mu_b = 0$):
\bea
& \int & \frac{dp}{2 \pi^2} p^2 C_{coll} E_i^{\alpha} =  -\left( \frac{n_i^2}{\bar{n}_i^2} - 1 \right) \nonumber \\
&  \times & \int d\Pi_{i} d\Pi_{j} d\Pi_{a} d\Pi_{b} (2 \pi)^4 \delta^{(4)}(p_i + p_j - p_a - p_b) |M|^2 \exp(-(E_i + E_j)/T) \times E_i^{\alpha}  \nonumber \\
 & = & -\left( n_i^2 - \bar{n}_i^2 \right) \langle \sigma v E_i^{\alpha} \rangle .
\eea
For $\alpha = 0$, the above equation is the well known contribution from thermal scatterings to the collision term. To estimate its value for $\alpha = 1$, we assume:
\be
\langle \sigma v E \rangle \simeq \langle \sigma v \rangle \langle E_i \rangle = \langle \sigma v \rangle \frac{\rho_i}{n_i} \label{eq:app}
\ee
where $\langle \;\; \rangle$ represents thermal average.
Thus:
\be
\int \frac{dp}{2 \pi^2} p^2 C_{coll} E_i^{\alpha}  = \left( \bar{n}_i^2 - n_i^2 \right) \left\{ \begin{array}{rl}  
\langle \sigma v \rangle & \mbox{, for $\alpha = 0$} \\
\langle \sigma v \rangle \frac{\rho_i}{n_i} &\mbox{, for $\alpha = 1$}
\end{array} \right. .
\label{eq:collfin}
\ee

\subsection{Decay Term}

Now we derive a simplified expression for the decay term, under approximations similar to the ones used in the last section.
 The decay term includes the contributions from particle decay and injection from other decays and is given 
by~\cite{kkn}:
\be
C_{dec}(p,t) = -\Gamma_i \frac{m_i}{E_i} F_i + BR(j \to i) \Gamma_j \frac{m_j}{p^2} \int_{m_j^2/4 p}^{\infty} dq F_j \left( \sqrt{(p+q)^2 -m_j^2},t \right) \label{eq:dec1}
\ee
where $BR(j \to i)$ is the branching ratio for $j \to i + X$ decay times the multiplicity of i particles in the final 
state\footnote{Eq.~(\ref{eq:dec1}) assumes a two body decay kinematics of the type $j \to i + X$, where $X$ can be equal to $i$, but
$m_i,m_X \ll m_j$. 
Decays with small mass splitings or three body decays are {\it not} described by Eq.~(\ref{eq:dec1}).}. 
Once again we consider the integral:
\bea
& & \int \frac{dp}{2 \pi^2} p^2 C_{dec}(p,t) E_i^{\alpha} =  -\Gamma_i \int \frac{dp}{2 \pi^2} p^2 \frac{m_i}{E_i} F_i(p,t) E_i^{\alpha} \nonumber \\
 & & + BR(j \to i) \Gamma_j m_j \int \frac{dp}{2 \pi^2} E_i^{\alpha} \int_{m_j^2/4 p}^{\infty} dq F_j \left( \sqrt{(p+q)^2 -m_j^2},t \right) \label{eq:dec2}
\eea
with $\alpha = 0 (1)$ for the contribution to the number (energy) density equation. 
The first term in Eq.~(\ref{eq:dec2}) gives:
\be
-\Gamma_i \int \frac{dp}{2 \pi^2} p^2 \frac{m_i}{E_i} F_i(p,t) E_i^{\alpha} = \left\{ \begin{array}{rl}
-\Gamma_i m_i n_i \langle \frac{1}{E_i} \rangle  & \mbox{, for $\alpha = 0$} \\
-\Gamma_i m_i n_i &\mbox{, for $\alpha = 1$}
\end{array} \right. .
\label{eq:dec1a}
\ee

The second term in Eq.~(\ref{eq:dec2}) can be simplified if we note that, for $m_j \gg m_i$, the $i$ particles injected
from $j$ decays are relativistic, so we have $E_i^{\alpha} \simeq p^{\alpha}$. 
The  integrals over $p$ and $q$ can be conveniently rewritten through the following change of variables:
\be
q \equiv \sqrt{P^2 + m_i^2} -p \mbox{ and } p \equiv \frac{m_i^2}{2 \left(\sqrt{P^2 + m_i^2} - P \cos\theta\right)} .
\ee
After performing the integration over $\cos\theta$, we obtain:
\be
\int \frac{dp}{2 \pi^2} p^{\alpha} \int_{m_j^2/4 p}^{\infty} dq F_j \left( \sqrt{(p+q)^2 -m_j^2},t \right) =  \left\{ \begin{array}{rl}
n_j \langle \frac{1}{E_j} \rangle & \mbox{, for $\alpha = 0$}  \\
n_j/2  & \mbox{, for $\alpha = 1$}
\end{array} \right. . 
\label{eq:dec1b}
\ee
Finally, replacing Eqs.~(\ref{eq:dec1a}) and (\ref{eq:dec1b}) in Eq.~(\ref{eq:dec2})
 and assuming $\langle 1/E \rangle \simeq 1 /\langle E \rangle = n/\rho$ (as in Eq.~(\ref{eq:app})), we have:
\be
\int \frac{dp}{2 \pi^2} p^2 C_{dec}(p,t) E_i^{\alpha} =   \left\{ \begin{array}{rl}
-\Gamma_i m_i n_i^2/\rho_i + BR(j \to i) \Gamma_j m_j n_j^2/\rho_j  & \mbox{, for $\alpha = 0$}  \\
-\Gamma_i m_i n_i + BR(j \to i) \Gamma_j m_j n_j/2  & \mbox{, for $\alpha = 1$}
\end{array} \right. .
\label{eq:decfin}
\ee

\subsection{Number and Energy Density Equations}

Using the results of Eqs.~(\ref{eq:collfin}) and (\ref{eq:decfin}) in the Boltzmann equations for $n_i$ and $\rho_i$ 
(Eq.~(\ref{eq:meqs})),
we obtain:
\bea
\Drv{n_i}{t} & + & 3H n_i =  \left( \bar{n}_i^2 - n_i^2 \right) \langle \sigma v \rangle -\Gamma_i m_i \frac{n_i^2}{\rho_i} + \sum_{j \neq i} BR(j \to i) \Gamma_j m_j \frac{n_j^2}{\rho_j} \label{nieq} \\
\Drv{\rho_i}{t} & + & 3H (\rho_i + P_i) = \left( \bar{n}_i^2 - n_i^2 \right) \langle \sigma v \rangle \frac{\rho_i}{n_i}  -\Gamma_i m_i n_i + \sum_{j \neq i} BR(j \to i) \Gamma_j \frac{m_j}{2} n_j  . \nonumber
\eea
It is convenient to use the above results to obtain a simpler equation for $\rho_i/n_i$:
\be
\Drv{\rho_i/n_i}{t} \equiv \Drv{R_i}{t} = -3 H \frac{P_i}{n_i} + \sum_{j \neq i} BR(j \to i) \Gamma_j m_j \frac{n_j}{n_i} \left( \frac{1}{2} - \frac{n_j}{\rho_j} \frac{\rho_i}{n_i} \right) .
\label{eq:Rieq}
\ee

Besides the above equations, it is useful to consider the evolution equation for entropy:
\be
S \equiv \frac{2 \pi^2}{45} g_{*S}(T) T^3 R^3
\ee
where $R$ is the scale factor. With the above definition we have~\cite{kt}:
\bea
\dot{S} & = &\left(\frac{2 \pi^2}{45} g_{*S}(T) \frac{1}{S}\right)^{1/3} R^4 \sum_{i} R(i\to X) \frac{1}{\gamma_i} \Gamma_i \rho_{i} \nonumber \\
\To \dot{S} & = &\frac{R^3}{T} \sum_{i} R(i\to X) \Gamma_i m_i n_{i} 
\label{Seq}
\eea
where $R(i\to X)$ is the fraction of energy injected in the thermal bath from $i$ decays.
 
Defining:
\be
x = \ln(R/R_0),\;\; N_i = \ln(n_i/s_0),\;\; {\rm and}\;\; N_S = \ln(S/S_0)
\ee
we can write Eqs.~(\ref{Seq}), (\ref{nieq}) and (\ref{eq:Rieq}) as:
\bea
N_S' & = & \frac{1}{HT} \sum_{i} R(i\to X) \Gamma_i m_i \exp[N_i + 3 x - N_S] \label{Seqb} \\
N_i' & = & - 3 - \frac{\Gamma_i}{H} \frac{m_i}{\rho_i/n_i} + \sum_{j\neq i} BR(j \to i) \frac{\Gamma_j}{H} \frac{m_j}{\rho_j/n_j} \frac{n_j}{n_i} +  \frac{\sigv_i}{H} n_i [\left(\frac{\bar{n}_i}{n_i}\right)^2 -1]  \\
R_i' & = & -3 \frac{P_i}{n_i} + \sum_{j \neq i} BR(j \to i) \frac{\Gamma_j}{H} m_j \frac{n_j}{n_i} \left( \frac{1}{2} - \frac{n_j}{\rho_j} \frac{\rho_i}{n_i} \right) \label{Nieq}
\eea
where $'=d/dx$.

The above equation for $N_i$ also applies for coherent oscillating fields, if we define:
\be
N_i = \ln(n_i/s_0),\;\; {\rm and}\;\; n_i \equiv \rho_i/m_i
\ee
so
\bea
N_i' & = & -3 - \frac{\Gamma_i}{H}  \nonumber \\
R_i'& = & 0 \label{Nico}
\eea
where we assume that the coherent oscillating component does not couple to any of the other fields.

Collecting Eqs.~(\ref{Seqb})-(\ref{Nieq}) and (\ref{Nico}), we have a closed set of first order differential equations:
\bi
\item Entropy:
\be
N_S' = \frac{1}{HT} \sum_{i} R(i\to X) \Gamma_i m_i \exp[N_i + 3 x - N_S] 
\label{NSeq}
\ee
\item Thermal fields:
\bea
N_i' & = & - 3 - \frac{\Gamma_i}{H} \frac{m_i}{\rho_i/n_i} + \sum_{j\neq i} BR(j \to i) \frac{\Gamma_j}{H} \frac{m_j}{\rho_j/n_j} \frac{n_j}{n_i} 
+  \frac{\sigv_i}{H} n_i [\left(\frac{\bar{n}_i}{n_i}\right)^2 -1] \nonumber \\
R_i' & = & -3 \frac{P_i}{n_i} + \sum_{j \neq i} BR(j \to i) \frac{\Gamma_j}{H} m_j \frac{n_j}{n_i} \left( \frac{1}{2} \label{eq:TPeq}
- \frac{n_j}{\rho_j} \frac{\rho_i}{n_i} \right)
\eea
\item Coherent Oscillating fields:
\bea
N_i' & = & -3 - \frac{\Gamma_i}{H} \nonumber \\
R_i' & = & 0 .
\label{eq:COeq}
\eea
\ei

As seen above, the equation for $R_i = \rho_i/n_i$ depends on $P_i/n_i$. A proper evaluation of this quantity
requires knowledge of the distribution $F_i(p,t)$. However, for relativistic (or massless) particles we have $P_i = \rho_i/3$,
 as seen from Eq.~(\ref{beqs}), whilst for particles at rest we have $P_i = 0$. 
Hence $F_i(p,t)$ is only required to evaluate the relativistic/non-relativistic transition, 
which corresponds to a relatively small part of the evolution history of particle $i$. 
Nonetheless, to model this transition we approximate $F_i$ by a thermal distribution and take $T_i, \mu_i \ll m_i$, 
where $T_i$ is
the temperature of the particle (which can be different from the thermal bath's). 
Under these approximations we have:
\be
\frac{P_i}{n_i} = T_i \;\; \mbox{ and } \;\; \frac{\rho_i}{n_i} = T_i \left[ \frac{K_1(m_i/T_i)}{K_2(m_i/T_i)} \frac{m_i}{T_i} + 3 \right] \label{eq:p1}
\ee
where $K_{1,2}$ are the modified Bessel functions. In particular, if $m_i/T_i \gg 1$:
\be
\frac{\rho_i}{n_i} \simeq T_i \left[\frac{3}{2} + \frac{m_i}{T_i}  + 3 \right] \To \frac{P_i}{n_i} = T_i = \frac{2 m_i}{3}\left( \frac{R_i}{m_i} -1 \right) .
\ee
As shown above, for a given value of $R_i = \rho_i/n_i$, Eq.~(\ref{eq:p1}) can be inverted to compute $T_i$ ($=P_i/n_i$):
\be
\frac{P_i}{n_i} = T_i(R_i) .
\ee
Since we are interested in the non-relativistic/relativistic transition, we can expand the above expression around $R_i/m_i = 1$,
so $P_i/n_i$ can be written as:
\be
\frac{P_i}{n_i} = \frac{2 m_i}{3}\left( \frac{R_i}{m_i} -1 \right) + m_i \sum_{n >1} a_n \left(\frac{R_i}{m_i} -1 \right)^n
\ee
where the coefficients $a_n$ can be numerically computed from Eq.~(\ref{eq:p1}). 
The above approximation should be valid for
$m_i/T_i \gtrsim 1$ (or $R_i \gtrsim m_i$). On the other hand, for $m_i/T_i \ll 1$ (or $R_i \gg m_i$), we have the 
relativistic regime, with $P_i/n_i = R_i/3$.
Therefore we can approximate the $P_i/n_i$ function for all values of $R_i$ by:
\be
\frac{P_i}{n_i} = \left\{ \begin{array}{rl}
& \frac{2 m_i}{3}\left( \frac{R_i}{m_i} -1 \right) + m_i \sum_{n >1} a_n \left(\frac{R_i}{m_i} -1 \right)^n  \mbox{ , for $R_i < \tilde{R}$} \\
& \frac{R_i}{3}  \mbox{ , for $R_i > \tilde{R}$} 
\end{array} \right. \label{Pfin}
\ee
where the coefficients $a_n$ are given by the numerical fit of Eq.~(\ref{eq:p1}) and 
$\tilde{R}$ is given by the matching of the two solutions.

Finally, to solve Eqs.~(\ref{NSeq})-(\ref{eq:COeq}) we need to compute $H$ according 
to Eq.~(\ref{H}), which requires knowledge of the energy densities for all
particles ($\rho_i$) and for the thermal bath ($\rho_R$). The former are directly obtained from $N_i$ and $R_i$, while the latter can be
 computed from $N_S$:
\be
T = \left(\frac{g_{*S}(T_R)}{g_{*S}(T)}\right)^{1/3} T_R \exp[N_S/3 -x] \To \rho_R = \frac{\pi^2}{30} g_{*}(T) T^4 .
\ee

Eqs.~(\ref{NSeq})-(\ref{eq:COeq}), with the auxiliary equations for $H$ (Eq.~(\ref{H})) and $P_i/n_i$ (Eq.~(\ref{Pfin})) 
 form a set of closed equations, which can be solved once the initial conditions for the number
density ($n_i$), energy density ($\rho_i$) and entropy ($S$) are given. For thermal fluids we assume:
\bea
n_i(T_R) & = & \left\{ 
\begin{array}{ll} 
0 & , \mbox{ if $\sigv_i \bar{n}_i/H|_{T=T_R} < 2$} \\
\bar{n}_i(T_R) & , \mbox{ if $\sigv_i \bar{n}_i/H|_{T=T_R} > 2$} 
\end{array} \right. \label{ni0TP} \\
\frac{\rho_i}{n_i}(T_R) & = & \frac{\bar{\rho}_i}{\bar{n}_i}(T_R)
\eea
where $\bar{\rho}_i$ is the equilibrium energy density (with zero chemical potential) for the particle $i$.
For coherent oscillating fluids, the initial condition is set at the beginning of oscillations:
\bea
n_i(T^{osc}_i) & = &\frac{\rho_i^{0}}{m_i(T^{osc}_i)} \\
\frac{\rho_i}{n_i}(T^{osc}_i) & = & m_i
\eea
where $T^{osc}_i$ is the oscillation temperature, given by $3H(T^{osc}_i) = m_i(T^{osc}_i)$ and $\rho_i^{0}$ the
initial energy density for oscillations.

Finally, the initial condition for the entropy $S$ is trivially obtained, once we assume a radiation dominated universe
at $T=T_R$:
\be
S(T_R) = \frac{2 \pi^2}{45} g_*(T_R) T_R^3 R_0^3 .
\ee

%

%
\end{document}